\documentclass[fleqn,usenatbib]{mnras}

\usepackage{newtxtext,newtxmath}

\usepackage[T1]{fontenc}

\DeclareRobustCommand{\VAN}[3]{#2}
\let\VANthebibliography\thebibliography
\def\thebibliography{\DeclareRobustCommand{\VAN}[3]{##3}\VANthebibliography}

\usepackage{graphicx}	
\usepackage{amsmath}	
\usepackage{amssymb}	

\usepackage{multirow}
\usepackage{caption}
\usepackage{subcaption}
\usepackage{setspace}
\usepackage{booktabs}

\title[Bayesian DL for radio galaxy classification]{Quantifying Uncertainty in Deep Learning Approaches to Radio Galaxy Classification}

\author[D. Mohan et al.]{
Devina Mohan,$^{1}$\thanks{E-mail: \url{ devina.mohan@postgrad.manchester.ac.uk}}
Anna M.~M.~Scaife,$^{1,2}$
Fiona Porter,$^{1}$
Mike Walmsley,$^{1}$
and
Micah Bowles$^{1}$
\\
$^{1}$Jodrell Bank Centre for Astrophysics, Department of Physics \& Astronomy, University of Manchester, M13 9PL, UK\\
$^{2}$Alan Turing Institute, 96 Euston Rd, London NW1 2DB, UK\\
}

\date{Accepted XXX. Received YYY; in original form ZZZ}

\pubyear{2015}

\begin{document}
\label{firstpage}
\pagerange{\pageref{firstpage}--\pageref{lastpage}}
\maketitle

\begin{abstract}
 In this work we use variational inference to quantify the degree of uncertainty in deep learning model predictions of radio galaxy classification. We show that the level of model posterior variance for individual test samples is correlated with human uncertainty when labelling radio galaxies. We explore the model performance and uncertainty calibration for different weight priors and suggest that a sparse prior produces more well-calibrated uncertainty estimates. Using the posterior distributions for individual weights, we demonstrate that we can prune 30\% of the fully-connected layer weights without significant loss of performance by removing the weights with the lowest signal-to-noise ratio. A larger degree of pruning can be achieved using a Fisher information based ranking, but both pruning methods affect the uncertainty calibration for Fanaroff-Riley type~I and type~II radio galaxies differently. Like other work in this field, we experience a cold posterior effect, whereby the posterior must be down-weighted to achieve good predictive performance. We examine whether adapting the cost function to accommodate model misspecification can compensate for this effect, but find that it does not make a significant difference. We also examine the effect of principled data augmentation and find that this improves upon the baseline but also does not compensate for the observed effect. We interpret this as the cold posterior effect being due to the overly effective curation of our training sample leading to likelihood misspecification, and raise this as a potential issue for Bayesian deep learning approaches to radio galaxy classification in future.
\end{abstract}




\section{Introduction}

A new generation of radio astronomy facilities around the world such as the Low-Frequency Array  \citep[LOFAR;][]{VanHaarlem2013}, the Murchison Widefield Array \citep[MWA;][]{Beardsley2019}, the MeerKAT telescope \citep{Jarvis2016}, and the Australian SKA Pathfinder (ASKAP) telescope \citep{Johnston2008} are generating large volumes of data. In order to extract scientific impact from these facilities on reasonable timescales, a natural solution has been to automate the data processing as far as possible and this has lead to the increased adoption of machine learning methodologies. 

In particular for new sky surveys, automated classification algorithms are being developed to replace the \emph{by eye} approaches that were possible historically. In radio astronomy specifically, studies looking at morphological classification using convolutional neural networks (CNNs) and deep learning have become increasingly common, especially with respect to the classification of radio galaxies. 

The Fanaroff-Riley (FR) classification of radio galaxies was introduced over four decades ago \citep{fr1974}, and has been widely adopted and applied to many catalogues since then. The morphological divide seen in this classification scheme has historically been explained primarily as a consequence of differing jet dynamics. Fanaroff-Riley type I (FRI) radio galaxies have jets that are disrupted at shorter distances from the central super-massive black hole host and are therefore centrally brightened, whilst Fanaroff-Riley type II (FRII) radio galaxies have jets that remain relativistic to large distances, resulting in bright termination shocks. These observed structural differences may be due to the intrinsic power in the jets, but will also be influenced by local environmental densities \cite{bicknell1995,kaiserbest}.   

Intrinsic and environmental effects are difficult to disentangle using radio luminosity alone as systematic differences in particle content, environmental effects and radiative losses make radio luminosity an unreliable proxy for jet power \citep{croston2018}. Hence the use of morphology is important for gaining a better physical understanding of the FR dichotomy, and of the full morphological diversity of the population, which in turn is useful for inferring the environmental impact on radio galaxy populations \citep{mingo2019revisiting}. It is hoped that the new generation of radio surveys, with improved resolution, sensitivity and dynamic range, will play a key part in finally answering this question.

From a deep learning perspective, the ground work for morphological classification in this field was done by \cite{aniyan2017classifying}, who used CNNs to classify FRI, FRII and bent-tail sources. This was followed by other works involving the use of deep learning in source classification \citep[e.g.][]{lukic2018,rgz,wu2018}. More recently, \cite{bowles2021attention} showed that an attention-gated CNNs could perform classification of radio galaxies with equivalent performance to other applications in the literature, but using $\sim$50\% fewer learnable parameters, \cite{e2cnn} showed that using group-equivariant convolutional layers that preserved the rotational and reflectional isometries of the Euclidean group resulted in improved overall model performance and stability of model confidence for radio galaxies at different orientations, and \cite{bastien2020} generated synthetic populations of radio galaxies using structured variational inference.

Applying deep learning to radio astronomy comes with unique challenges. Unlike terrestrial labelled datasets such as MNIST, which contains $\sim$ 70,000 images, and ImageNet, which contains 14 million images, there is a dearth of labelled data in radio astronomy. The largest labelled datasets for Fanaroff-Riley classification contain of order $10^3$ labelled images\footnote{We note that the Radio Galaxy Zoo catalogue \citep{rgz} contains of order $10^4$ objects, but these are not labelled by morphological type.}. For deep learning applications, this creates the need to augment datasets. However, this augmentation can lead to any biases associated with these small datasets being propagated into the larger augmented datasets used to train deep learning models and hence into any analysis that uses the outputs of those models.

Another challenge is that of artefacts, misclassifed objects and ambiguity arising from how the morphologies in these datasets are defined. Underestimation and miscalibration of uncertainties associated with model outcomes for data samples that are peripheral to the main data mass are well documented in the machine learning literature, see e.g. \cite{guo2017}, and it has been demonstrated that out-of-distribution data points will be misclassified with arbitrarily high precision by standard neural networks \citep{hein2019}. 

To provide uncertainties on model outputs, probabilistic methods such as Bayesian (and approximately Bayesian) neural networks are required \citep{mackay1992a,mackay1992b}. When properly calibrated, the uncertainty estimates from these approaches can serve as a diagnostic tool to mitigate the effect of increasingly distant data points and out-of-distribution examples. However, with the exception of \cite{e2cnn}, to date there has been little work done on understanding the degree of confidence with which CNN models predict the class of individual radio galaxies. In modern radio astronomy, where astrophysical analysis is driven by population analyses, quantifying the confidence with which each object is assigned to a particular classification is crucial for understanding the propagation of uncertainties within that analysis. 

In this work we use variational inference (VI) to implement a fully Bayesian CNN and quantify the degree of uncertainty in deep learning predictions of radio galaxy classifications. This differs from the approach of \cite{e2cnn} who used dropout as a Bayesian approximation to estimate model confidence \citep{gal2016}. They studied one specific aspect of the model performance (variation with sample orientation), and as such it is not directly comparable to this work.
We compare the variance of our posterior predictions to qualifications present in our test data that indicate the level of human confidence in assigning a classification label and show that model uncertainty is correlated with human uncertainty. We also investigate a number of the challenges that face the systematic use of Bayesian deep learning from the perspective of radio astronomy.



The structure of the paper is as follows: in Section~\ref{sec:vi_intro} we introduce the variational inference method and its application to neural networks; in Section~\ref{sec:uncertainty} we describe how different measures of uncertainty can be recovered from the learned variational posteriors using this approach; and in Section~\ref{sec:data} we describe the dataset being used in this work. In Section~\ref{sec:model} we introduce the convolutional neural network that forms the primary model for this work, as well as how it is trained; in Section~\ref{sec:results} we describe the results of that training in terms of model performance and uncertainty quantification in the context of the specific radio galaxy classification problem being addressed, as well as the wider machine learning literature; in Section~\ref{sec:masegosa} we discuss the cold posterior effect and hypotheses for mitigating it. In Section~\ref{sec:conclusion} we draw our conclusions. 
\section{Variational Inference for Deep Learning}
\label{sec:vi_intro}

The notion of "noisy weights" that can adapt during training was first proposed by \cite{hinton1993noisy} to reduce the amount of information in network weights and prevent overfitting in neural networks. \citet{practicalvi} developed a stochastic variational inference (SVI) method by applying stochastic gradient descent to VI using biased estimates of gradients. SVI allows VI to scale to large datasets by taking advantage of mini-batching and \cite {practicalvi} considered various choices of standard prior and posterior distributions such as the Delta function, Gaussian and Laplace distributions.
\citet{blundell} built on this work and proposed the \emph{Bayes by Backprop} algorithm, which combines stochastic VI with the reparameterization trick \citep{kingma2015variational} to overcome the problems encountered while using backpropagation with SVI. Using this algorithm, one can calculate unbiased estimates of the gradients and use any tractable probability distribution to represent uncertainties in the weights. 

To set up the problem of Bayesian inference, we consider a set of observations, $\mathbf{x}$, and a set of hypotheses, $\mathbf{z}$. For instance, for a neural network $\mathbf{z}$ are the parameters of the model.

Bayes rule allows us to compute the posterior distribution, $p(z|x)$:
\begin{equation}
    p(z|x) = \frac{ p(z,x)}{p(x)} = \frac{p(x|z) \, p(z)}{p(x)} , 
    \label{eq:bayes}
\end{equation}
where $p(x|z)$ is the \emph{likelihood} of the data given the hypothesis and quantifies how well the hypothesis fits to the data; $p(z)$ is the prior distribution; and the denominator, which is called the \emph{evidence}, is the marginalised probability of the observations.

To compute the evidence, we need a solution to the following integral:
\begin{equation}
    p(x) = \int p(z,x) \, {\rm d}z = \int p(x|z) \, p(z) \, {\rm d}z,
    \label{eq:evidence}
\end{equation}
which is obtained by integrating over all possible values of $z$. This integral is often intractable either because there is no closed form solution available or the computation is exponential in time because it requires evaluating the integral for all possible values of z, which is very high dimensional. This in turn makes the posterior, $p(z|x)$, intractable.

\subsection{The Variational Inference Cost Function}

In variational inference, a parameterised probability distribution, $q(z)$, is defined as a variational approximation to the true posterior, $p(z|x)$. The family of probability distributions, $\mathbb{D}$, defines the complexity of the solution that can be modelled. For instance, a family of Gaussians parameterised by mean, ${\mu}$, and variance, ${\sigma^2}$, may be defined. The goal of VI is to find the selection of parameters which most closely approximates the exact posterior. 

The difference between the variational posterior, $q(z)$, and the exact posterior, $p(z|x)$, is typically measured using the KL divergence \citep{kullback1951information}. Therefore, to find the optimal variational density from the family of densities, the following optimisation problem is solved:
\begin{equation}
    q^*(z) =  \operatorname*{arg\,min}_{q \in \mathbb{D} } {\rm KL}[q(z) || p(z|x)] .
    \label{eq:min kl}
\end{equation}
If we expand above equation using the formula for KL divergence:
\begin{equation}
    {\rm KL}[a(x)||b(x)] = \int a(x) \log \frac{a(x)}{b(x)} \, {\rm d}x,
    \label{eq:kl}
\end{equation}
we get:
\begin{eqnarray}
\nonumber     q^*(z) &=& \operatorname*{arg\,min}_{q \in \mathbb{D} } \int q(z) \log \frac{q(z)}{p(z|x)} \, {\rm d}z\\
\nonumber    & = &\operatorname*{arg\,min}_{q \in \mathbb{D} } \int q(z) \log q(z) \, {\rm d}z - \int q(z) \log p(z|x) \, {\rm d}z \\ 
\nonumber    & = &\operatorname*{arg\,min}_{q \in \mathbb{D} } \mathbb{E}_{q(z)} [\log q(z)] - \mathbb{E}_{q(z)} [\log p(z|x)] \\
\nonumber    & = &\operatorname*{arg\,min}_{q \in \mathbb{D} } \mathbb{E}_{q(z)} [\log q(z)] - \mathbb{E}_{q(z)} [\log p(z,x)] + \log p(x),\\
    \label{eq:min kl 1}
\end{eqnarray}
where the expectation value is calculated according to:
\begin{equation}
    \mathbb{E}_{b(x)}[a(x)] = \int a(x)b(x)\, \rm{d}x .
    \label{eq:expectation}
\end{equation}
We can see the dependence of minimising the KL divergence on the intractable integral $p(x)$ in Equation~\ref{eq:min kl 1}. Therefore, to find the optimal variational density $q^*(z)$, the above equation is only minimised up to an additive constant:

\begin{equation}
    q^*(z)=\operatorname*{arg\,min}_{q \in \mathbb{D} } \mathbb{E}_{q(z)} [\log q(z)] - \mathbb{E}_{q(z)} [\log p(z,x)] .
    \label{eq:min kl2}
\end{equation}

Minimising the above function is equivalent to maximising the following objective function, which is formulated in the literature as the Evidence Lower Bound (ELBO) or variational free energy \citep{saul1996mean, neal1998view}:
\begin{eqnarray}
\nonumber    {\rm ELBO}(q) & = &  \mathbb{E}_{q(z)} \log p(z,x) - \mathbb{E}_{q(z)} [\log q(z)] \\
\nonumber     & = & \mathbb{E}_{q(z)} \log p(x|z) + \mathbb{E}_{q(z)}\log p(z) - \mathbb{E}_{q(z)} [\log q(z)] \\
     & = & \mathbb{E}_{q(z)} \log p(x|z) - {\rm KL}(q(z)||p(z)).
    \label{eq:elbo}
\end{eqnarray}
    
The name ELBO stems from the fact that the log evidence is bounded by this function such that: $\log p(x) \geq \rm{ELBO}$. Consequently, variational inference reduces Bayesian inference to an optimisation problem, which can then be solved by standard deep learning optimisation algorithms such as SGD and Adam.

\subsection{VI for Neural Networks}

More specifically, for neural networks, a family of distributions with parameters, $\theta$, is posited over the network parameters\footnote{Here we refer only to the weights, $w$, of the network for simplicity, but the equations are applicable to the biases as well.} to define a variational approximation to the posterior, $q(w|\theta)$. Following Equation~\ref{eq:min kl}, one can find a member of the family that is closest to the true Bayesian posterior, $P(w|D)$, by minimising the following objective function:
\begin{equation}
     \theta^* = \operatorname*{arg\,min}_{\theta} {\rm KL} [q(w|\theta)||P(w|D)] ,
\end{equation}
where $D$ denotes the training data. Using the formula for KL divergence given in Equation~\ref{eq:kl} we can write
\begin{eqnarray}
\nonumber  \theta^*  & = & \operatorname*{arg\,min}_{\theta} \int q(w|\theta) \log \frac{q(w|\theta)}{P(w|D)}\,{\rm d}w \\
\nonumber    & = & \operatorname*{arg\,min}_{\theta} \int q(w|\theta) \log \frac{q(w|\theta)}{P(w)P(D|w)}\,{\rm d}w \\
\nonumber    & = &\operatorname*{arg\,min}_{\theta} \int q(w|\theta) [\log q(w|\theta) - \log P(w) \\
\label{eq:cost_blundell}    && \hspace{12em} - \log P(D|w)]\,{\rm d}w \\
\nonumber    & = &\operatorname*{arg\,min}_{\theta} \int q(w|\theta) [\log \frac{q(w|\theta)}{P(w)} {\rm d}w \\
\nonumber && \hspace{10em} - \int q(w|\theta) \log P(D|w)]\,{\rm d}w \\
\label{eq:cost}    & = &\operatorname*{arg\,min}_{\theta} {\rm KL} [q(w|\theta)|P(w)] - \mathbb{E}_{q(w|\theta)} [\log P(D|w)] .
\end{eqnarray}

The cost function shown in Equation~\ref{eq:cost} is composed of two components: the first term is a \emph{complexity cost} which depends on the prior over the weights, $P(w)$, and the second is a \emph{likelihood cost} which depends on the data and describes how well the model fits to the data. The cost function also has a minimum description length interpretation according to which the best model is the one that minimises the cost of describing the model and the misfit between the model and the data to a receiver \citep{hinton1993noisy, practicalvi}.

More practially, the cost function used by \citet{blundell} is given by Equation~\ref{eq:cost_blundell}, which can be simplified as follows:
\begin{eqnarray}
\nonumber     \mathcal{F} (D,\theta) & = & \int q(w|\theta) [\log q(w|\theta) - \log P(w) \\
\label{eq:int_cost_blundell}     && \hspace{10em} - \log P(D|w)]\, {\rm d}w \\
    & = &\int q(w|\theta) f(w,\theta) \,{\rm d}w \\
    & = & \mathbb{E}_{q(w|\theta)} [f(w,\theta)] .
\end{eqnarray}

The cost, $\mathcal{F}$, is an expectation of the function, $f(w,\theta)$, with respect to the variational posterior, $q(w|\theta)$. In order to optimise the cost function, we need to calculate its gradient with respect to the variational parameters, $\theta$. To make $\mathcal{F}(D,\theta)$ differentiable, one must first employ the reparameterization trick \citep{kingma2013auto, kingma2015variational} to calculate samples from the variational posterior, $q(w|\theta)$, that are differentiable and then use Monte Carlo (MC) estimates of the gradients to approximate the cost function. 

The reparameterization trick makes use of the change of variables technique to map between probability densities of random variables. One can sample a random deviate from a known probability distribution and map it to a sample from the variational posterior through a differentiable deterministic function. This allows us to reparameterize samples from the variational posterior:
\begin{equation}
    w \sim q(w|\theta),
\end{equation}
through a differentiable deterministic function:
\begin{equation}
    w= t(\epsilon, \theta),
\end{equation}
where $\epsilon$ is a random deviate sampled from the distribution $\epsilon \sim q(\epsilon)$, such that:
\begin{equation}
     q(w|\theta) \, {\rm d}w = q(\epsilon) {\rm d} \epsilon . 
\end{equation}

By taking Monte Carlo samples from the variational posterior:
\begin{equation}
    w^{(i)} \sim q(w^{(i)}|\theta)  ,  
\end{equation}
Equation~\ref{eq:int_cost_blundell} becomes: 
\begin{equation}
    F(D, \theta) \approx \sum_{i=1}^n \log q (w^{(i)}|\theta) - \log P(w^{(i)}) - \log P(D| w^{(i)}),
    \label{eq:blundell_mccost}
\end{equation}
and this approach is typically referred to in the literature as \emph{Bayes by Backprop} (BBB).

\subsubsection{Mini-Batching}

To take advantage of mini-batch optimisation for Bayes by Backprop, a weighted complexity cost is used \citep{practicalvi}. This is because the likelihood cost is calculated for each mini-batch to update the weights when the model sees new data, whereas the complexity cost, which involves calculating the prior and posterior over the weights of the entire network, should be calculated only once per epoch because it is independent of data.


The simplest weighting factor one could use is the number of batches and the cost function for the $i^{{\rm th}}$ mini-batch in Equation~\ref{eq:cost} becomes:
%
\begin{equation}
    F_i(D_i, \theta) = \frac{1}{M} {\rm KL} [q(w|\theta]||P(w)] - \mathbb{E}_{q(w|\theta)} [\log P(D_i|w)] ,
    \label{eq:minibatch_cost_blundell}
\end{equation}
where $M = N_{{\rm batches}}$.

Several published results have reported a \emph{cold posterior effect} which involves further down-weighting of the complexity cost \citep[e.g.][]{wenzel2020}. This effect is discussed in more detail in Sections~\ref{sec:cold}~\&~\ref{sec:masegosa}.

\subsection{Variational Posteriors} 

The reparameterization trick allows us to use a variety of family densities for the variational distribution. \citet{kingma2013auto} give some examples of $q(w|\theta)$ for which the reparameterization trick can be applied. These include any tractable family of densities such as the Exponential, Logistic, Cauchy distributions; and any location-scale family such as Gaussian, Laplace, or Uniform densities can be used with the function: $t(.) = {\rm location} + {\rm scale} \cdot  \epsilon$.

For example, if we consider the case where the variational posterior, $q(w|\theta)$, is parameterised by a family of Gaussians then the variational parameters will be $\theta = (\mu, \rho) $, where the standard deviation, $\sigma$, is parameterised as $\log (1+\exp(\rho))$, so that it remains positive. To obtain a posterior sample, $w$, of the weight from the variational posterior one must sample $\epsilon$ from a unitary Gaussian, $\epsilon \sim \mathcal{N}(0, 1)$, and then map $\epsilon$ to a Gaussian distribution, $\mathcal{N}(\mu, \sigma)$, through the function:
\begin{equation}
    t(\epsilon, \mu, \rho) = w = \mu + \epsilon . \log (1 + \exp(\rho)),
\end{equation}
i.e., scale the random deviate by the standard deviation and shift it by the mean of the variational posterior. 

Following \cite{blundell}, we can then calculate the gradient of the cost function with respect to the variational parameters $ \theta = (\mu, \rho)$ using the standard optimisation algorithms that are used with neural networks.
%
%
%
%
%

\subsection{Priors}
\label{sec:priors}
Priors reflect our beliefs about the distribution of weights before the model has seen any data. The simplest prior we could use is a Gaussian prior:
\begin{equation}
    P(w) = \prod_{j} \mathcal{N} \, (w_j|0,\sigma) ,
\end{equation}
which is often the default prior used with Bayesian neural networks. 

To allow for a wide range of weight values to be learned, \cite{blundell} suggest the use of a ``spike-and-slab'' Gaussian Mixture Model (GMM) prior defined over all the  weights in the network such that
\begin{equation}
    P(w) =  \prod_{j} \pi \, \mathcal{N}(w_j|0,\sigma_1) + (1-\pi) \mathcal{N}(w_j|0,\sigma_2),
\end{equation}
where $\sigma_1 > \sigma_2 $ and $\sigma_2 << 1$. The weight of each component in the mixture is defined by $\pi$. These parameters are chosen by comparing the model performance on the validation set in the same manner as for other hyperparameters of the model.

In this work we also consider a Laplace prior which is parameterised by a location parameter, $\mu$, and a scale parameter, $b$, and a Laplace Mixture Model (LMM) prior with two mixture components weighted by $\pi$, similar to in form to the definition of the GMM prior.

Some regularisation techniques used with point-estimate neural networks have theoretical justifications using Bayesian inference. For instance, it can be shown that \textit{Maximum a Posteriori} (MAP) estimation of neural networks with some priors is equivalent to regularisation \citep{jospin2020hands}. For example, using a Gaussian prior over the weights is equivalent to weight decay regularisation, whereas using a Laplace prior induces L1 regularisation. \citet{gal2015bayesian} showed that dropout can also be considered an approximation to variational inference, where the variational family is a Bernoulli distribution. 

\subsection{Bayesian Convolutional Neural Networks}
\label{sec:bayesian_cnn}

The Bayes by Backprop algorithm can be extended for convolutional neural networks by sampling the weights from a variational distribution defined over the shared weights of the convolutional kernels. This is followed by fully-connected layers that have weights with a variational distribution defined over them. For simplicity, our implementation differs from that proposed by \cite{shridhar2019}, in which work the activations of each convolutional layer were sampled instead of the weights in order to accelerate convergence. 


\subsection{Posterior Predictive Distribution}


After the variational posterior distribution has been learned by optimising the ELBO function, we can use it to predict the labels of new observations, $D^*$, using the posterior predictive distribution, $ q(D^*|D)$ (here we have dropped the $q^*$ notation for clarity). The variational posterior distribution conditioned on training data $D$, $q(w|D)$, can be used to calculate this distribution by integrating out the variational parameters: 
\begin{eqnarray}
\nonumber    q(D^*|D) & = & \int q(D^*,w | D) \, {\rm d} w \\
\nonumber    & = & \int q(D^*|w, D) q(w|D) \, {\rm d} w \\
    & = & \int q(D^*|w) q(w|D) \, {\rm d} w , 
    \label{eq:wtavg}
\end{eqnarray}
where $q(D^*|w, D) = p(D^*|w)$ because for a given $w$, all data are conditionally independent (i.i.d. assumptions).

From Equation~\ref{eq:wtavg}, we see how the posterior predictive distribution is an average of all possible variational parameters weighted by their posterior probability. 

\noindent
The posterior predictive distribution can be estimated using MC samples as follows:
\begin{enumerate}
    \item Sample variational parameters from the variational posterior distribution conditioned on data $D$: $w^{(i)} \sim q(w|D)$.
    \item Sample prediction $D^{* (i)}$ from $q(D^*|w^{(i)})$.
    \item Repeat steps (i) and (ii) to construct an approximation to $q(D^*|D)$ using $N$ samples such that:
    \begin{eqnarray}
         q(D^*|D) & = & \mathbb{E}_{q(w|D)} q(D^*|w) \\
         & = & \mathbb{E}_{w^{(i)} \sim q(w|D)} q(D^*|w^{(i)}) \\
         & \approx &  \frac{1}{N}\sum_{i=1}^N q(D^*|w^{(i)}).
         \label{eq:ppd_mc_samples}
    \end{eqnarray}
\end{enumerate}

Thus BBB can be used to construct an approximate posterior predictive distribution, which can further be used to estimate uncertainties. 

\section{Uncertainty Quantification}
\label{sec:uncertainty}

The sources of uncertainty in the predictions of neural network models can broadly be divided into two categories: epistemic and aleatoric \citep{gal2016uncertainty, uq2021review}. Epistemic uncertainty quantifies how uncertain the model is in its predictions and this can be reduced with more data. Aleatoric uncertainty on the other hand represents the uncertainty inherent in the data and cannot be reduced. Uncertainty inherent in the input data along with model uncertainty is propagated to the output, which gives us predictive uncertainty \citep{uq2021review}. BBB allows us to capture model uncertainty by defining distributions over model parameters. 


Using Monte Carlo (MC) samples obtained from the posterior predictive distribution, one can obtain $N$ Softmax probabilities for each class, $c$, in the dataset. 
Adapting Equation~\ref{eq:ppd_mc_samples} for our supervised classification setting, we can recover $N$ class-wise Softmax probabilities as follows:
\begin{equation}
    q(y|x, D) = \frac{1}{N}\sum_{i=1}^N q(y=c|x, w^{(i)}) ,
    \label{eq:softmax_probs}
\end{equation}
where $(x,y)$ are samples from the test set and $D$ is the training data. 
Using these samples, we can quantify the uncertainties in the predictions using the metrics defined in the following subsections.

\subsection{Predictive Entropy}
The predictive entropy is the sum of epistemic and aleatoric uncertainties. It is a measure of the average amount of information inherent in the distribution and is defined as:
\begin{equation}
     \mathbb{H}(y|x, D) = - \sum_c q(y=c|x, w) \log q(y=c|x, w) ,
\end{equation}
which can be approximated using MC samples as:
\begin{equation}
    \mathbb{H}(y|x, D) = - \sum_c \left (\frac{1}{N} \sum_{i=1}^N q(y=c|x, w^{(i)}) \right) \log \left (\frac{1}{N} \sum_{i=1}^N q(y=c|x, w^{(i)}) \right)
    \label{eq:pred_entropy}
\end{equation}
following \cite{gal2016uncertainty}.

We use the natural logarithm for all the equations described in this section and the values are reported in nats, which is the natural unit of information. The entropy thus attains a maximum value of $\sim 0.693$ nats, when the predictive entropy is maximum and a minimum value close to zero.

\subsection{Mutual Information}
We use mutual information to quantify epistemic uncertainty. Mutual information is closely related to the entropy and can be calculated as follows:
\begin{equation}
    \mathbb{I}(y, w|x, D) = \mathbb{H} (y|x, D)
    - \mathbb{E}_{q(w|D)} [\mathbb{H}(y|x, w)],
\end{equation}
which can be approximated as:
\begin{equation}
    \nonumber \mathbb{I}(y, w|x, D) =
\end{equation}
\begin{eqnarray}
\nonumber    &&- \sum_c \left ( \frac{1}{N} \sum_{i=1}^N q(y=c|x, w^{(i)}) \right ) \log \left (\frac{1}{N} \sum_{i=1}^N q(y=c|x, w^{(i)}) \right) \\
\label{eq:mi}    & & + \frac{1}{N} \sum_{c, N} q(y=c|x, w^{(i)}) \log q(y=c|x, w^{(i)}) ,
\end{eqnarray}
following \cite{gal2016uncertainty}.


\subsection{Average Entropy}

\cite{mukhoti2021deterministic} demonstrate that for classical NNs, the entropy of a single pass can be used to quantify aleatoric uncertainty for in-distribution data samples. Here we extend that definition to our Bayesian NN and take the average entropy for a single input using $N$ MC samples to capture the aleatoric uncertainty associated with the data point, such that: 
\begin{equation}
    \mathbb{E}_{q(w|D)} \mathbb{H}(y|x,w) = - \frac{1}{N} \sum_{c, N} q(y=c|x, w^{(i)}) \log q(y=c|x, w^{(i)}).
    \label{eq:avg_entropy}
\end{equation}

It can be seen from Equation~\ref{eq:mi} and Equation~\ref{eq:avg_entropy} that the predictive uncertainty in Equation~\ref{eq:pred_entropy} is a sum of epistemic uncertainty and aleatoric uncertainty.

\subsection{Overlap Index}

In addition to the uncertainty metrics described above, we also define two overlap indices: $\eta_{\rm soft}$, to quantify how much the distributions of predicted Softmax values for the two classes overlap; and $\eta_{\rm logits}$, to quantify how much the distributions of logits for the two classes overlap\footnote{Logits are the unnormalised outputs of the network, i.e., the outputs of the final layer before the Softmax function is applied.}. A higher degree of overlap indicates a higher level of predictive uncertainty. The overlap parameters have contributions from both epistemic and aleatoric uncertainties.


We calculate distribution free overlap indices \citep{pastore2019measuring, e2cnn} by first estimating the local density at a location $z$ using a Gaussian kernel density estimator separately for each class such that
    \begin{eqnarray}
        f_{c_1}(z) & = & \frac{1}{N} \sum_{i=1}^N \frac{1}{\beta \sqrt{2 \pi}} ~ {\rm e}^{-(z- c_1^N)^2/2 \beta^2} ~ ,\\
        f_{c_2}(z) & = & \frac{1}{N} \sum_{i=1}^N \frac{1}{\beta \sqrt{2 \pi}} ~ {\rm e}^{-(z- c_2^N)^2/2 \beta^2} ~ ,
    \end{eqnarray}
    where $\beta = 0.1$ and $c_1$, $c_2$ are the softmax/logit values of the two classes in the dataset. The overlap index, $\eta$, can then be calculated using those local densities:
    \begin{equation}
        \eta = \sum_{i=1}^{M_z} {\rm min} [f_{c_1}(z_i), f_{c_2}(z_i)] \, \delta z \, ,
        \label{eq:eta}
    \end{equation}  
    where ${M_z}$ defines the step size of $z$ such that $\{z_i\}_{i=1}^{M_z}$ ranges from zero to one in $M_z$ steps.

\begin{figure*}
    \centerline{\includegraphics[width=\textwidth]{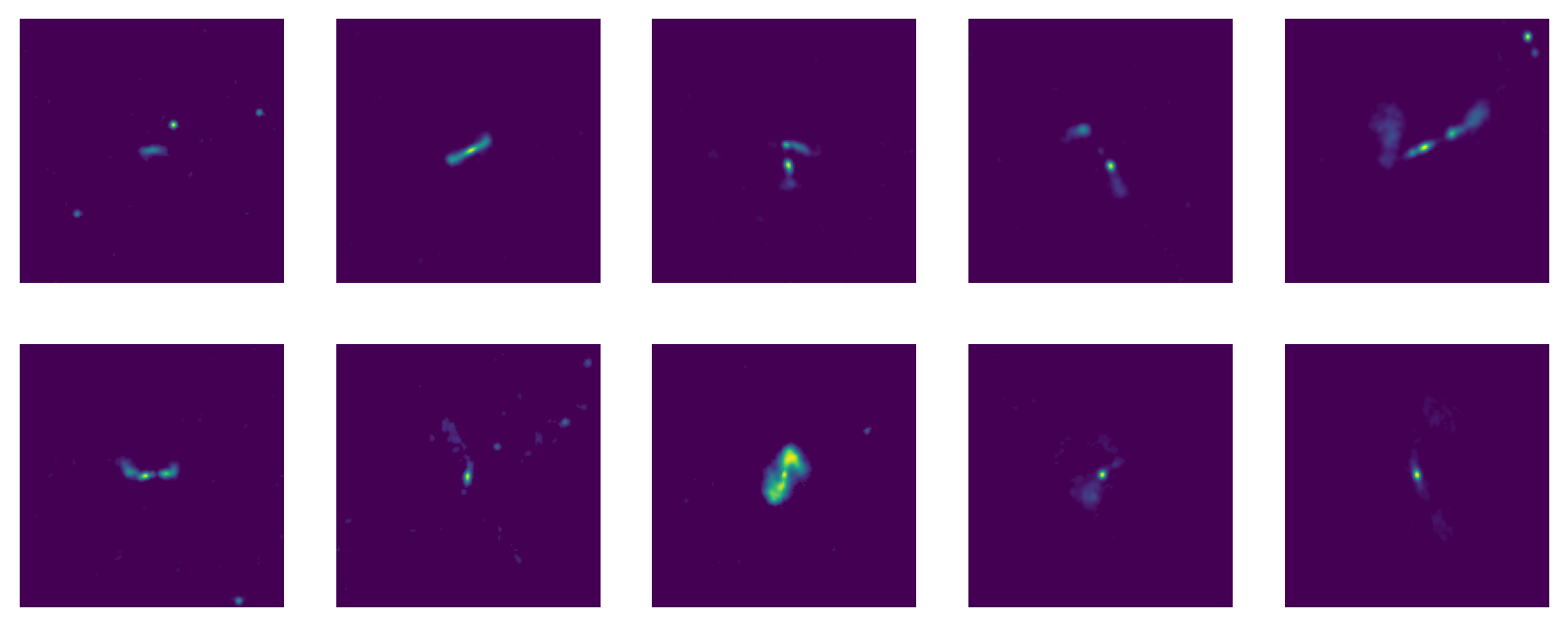}}
    \centerline{(a) Examples of FRI galaxies from the MiraBest Confident subset}
	\centerline{\includegraphics[width=\textwidth]{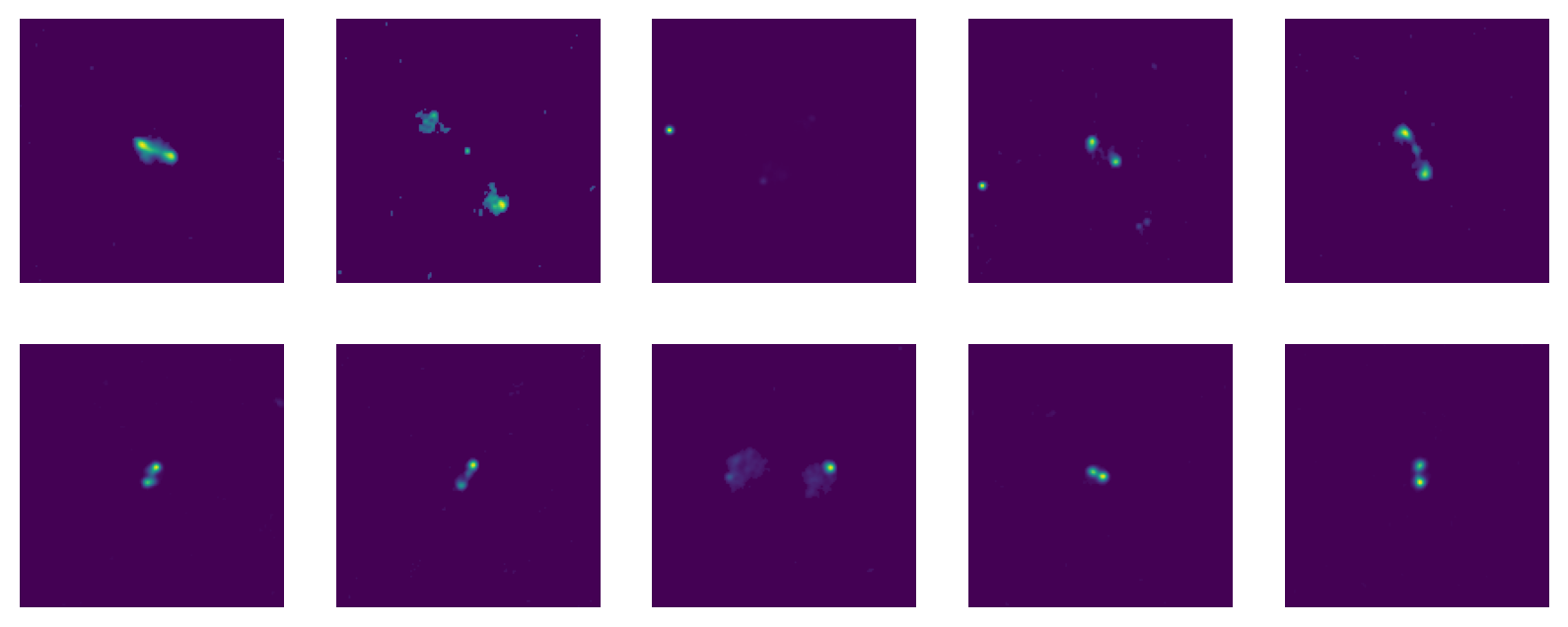}}
	\centerline{(b) Examples of FRII galaxies from the MiraBest Confident subset}
    \caption[Examples of (a) FRI and (b) FRII galaxies from the MiraBest Confident subset]{Examples of (a) FRI and (b) FRII galaxies from the MiraBest Confident subset}
	\label{fig:Mirabest Confident}
\end{figure*}

\subsection{Uncertainty Calibration}
\label{sec:calibration}


Bayesian neural networks allow us to obtain uncertainty estimates on model predictions, but these estimates have often been shown to be poorly calibrated due to the use of approximate inference methods and model misspecification \citep{foong2019expressiveness, krishnan2020improving}. A model is considered to be well-calibrated if the degree of uncertainty is correlated with the accuracy, i.e. low uncertainty predictions are more likely to be classified correctly and high uncertainty predictions are more likely to be misclassified. Therefore, when comparing different models one must also take the calibration of uncertainty metrics into account, in addition to the overall accuracy of a model. 

Following \cite{laves2019well}, we evaluate the calibration of each of our uncertainty metrics using the class-wise expected Uncertainty Calibration Error (cUCE). The UCE is a more generalised form of the widely used Expectation Calibration Error \citep[ECE;][]{guo2017calibration}; where the ECE evaluates the calibration of Softmax probabilities from classical point-wise models, the UCE evaluates the calibration of more general uncertainty metrics from probabilistic models, such as those described earlier in this section. \cite{laves2019well} define the cUCE value as an average of the UCE values obtained for each class, $c$, in the dataset:
\begin{equation}
    {\rm cUCE} = \frac{1}{C}\sum_{c=1}^C {\rm UCE}(c),
\end{equation}
where the UCE is a weighted average of the difference between fractional error and uncertainty calculated for the output of the model when binned into $M$ bins of equal width for a particular uncertainty metric:
\begin{equation}
    {\rm UCE} = \sum_{m=1}^M \frac{|B(m)|}{n}| {\rm err}(B_m) - {\rm uncert}(B_m)|.
\end{equation}
Here $B_m$ is the set of data in a particular bin, $n$ is the total number of data points and ${\rm uncert}(B_m)$ is the average value of a given uncertainty metric for those data points:
%
%
\begin{equation}
    {\rm uncert}(B_m) = \frac{1}{|B_m|} \sum_{i \in B_m} {\rm uncert_i} ~ ,
\end{equation}
where ${\rm uncert_i}$ can be calculated using Equation~\ref{eq:pred_entropy}, \ref{eq:mi} or \ref{eq:avg_entropy}, followed by minmax-normalisation to bring values into the range 0 to 1. 

Equation 16 of \cite{laves2019well} defines the average fractional error in bin $B_m$ to be:
\begin{equation}
 {\rm err}(B_m) = \frac{1}{|B_m|} \sum_{i \in B_m} {\rm err}_i~,
\end{equation}
where ${\rm err}_i$ is the contribution to this error from an individual data point, defined as:
\begin{equation}
 {\rm err}_i = \mathbf{1} (\hat{y}_i \neq y)~~~~~\forall~~i \in B_m~.
\end{equation}
Here we redefine ${\rm err}_i$ to be the average error obtained for an individual data sample, such that
\begin{equation}
 {\rm err}_i = \frac{1}{N} \sum_{j=1}^N \mathbf{1} (\hat{y}_{ij} \neq y)~~~~~\forall~~i \in B_m~,
\end{equation}
where $N$ is the number of samples drawn from the posterior predictive distribution. For a binary classification problem, the fractional error from an untrained model is expected to converge towards a value of 0.5, i.e. 50\% accuracy. Consequently we normalise the average fractional error per bin, ${\rm err}(B_m)$, by a factor of two in order to scale the values from 0 to 1.

\begin{table}
\centering
\caption[Short table caption.]{Three digit identifiers for sources in \citet{Miraghaei2017TheEnvironment}.}
	\begin{tabular}{llll}
    Digit 1 & Digit 2 & Digit 3 \\
    
    \hline
    
    1: FRI & 0: Confident & 0: Standard \\
    2: FRII & 1: Uncertain & 1: Double Double \\
    3: Hybrid &         & 2: Wide Angle Tail \\ 
    4: Unclassifiable & & 3: Diffuse \\
    &  &                   4: Head Tail \\ 
	\end{tabular}
   \label{tab:digits_mirabest}
\end{table}

\begin{table}

\centering
\caption[MiraBest Class-wise Composition]{MiraBest Class-wise Composition.}
	\begin{tabular}{cccc}
	\hline
    Class & Confidence & No. \\
    \hline
    \multirow{2}{2em}{FRI}  & Confident & 397\\
     & Uncertain & 194\\
    \hline
    \multirow{2}{2em}{FRII}  & Confident & 436\\
     & Uncertain & 195\\
    \hline
    \multirow{2}{2em}{Hybrid}  & Confident & 19\\
     & Uncertain & 15\\
    \hline
    
	\end{tabular}
   \label{tab:mb_classes}
\end{table}

\section{Data}
\label{sec:data}

Radio galaxies are a sub-class of active galactic nuclei (AGN). These galaxies are characterised by large scale jets and lobes which can extend up to mega-parsec distances from the central black hole and are observed in the radio spectrum. 
\citet{fr1974} proposed a classification of such extended radio sources based on the ratio of the distance between the highest surface brightness regions on either side of the galaxy to the total extent of the radio source, $R_{{\rm FR}}$. Based on a threshold ratio of 0.5, the galaxies were classified into two classes as follows: if $R_{{\rm FR}} < 0.5$, the source was classified into Class I (FRI; edge-darkened), and if $R_{{\rm FR}} > 0.5$ it was classified into Class II (FRII; edge-brightened). Over the years, several other morphologies such as bent-tail \citep{rudnick1976, odea1985owen}, hybrid \citep{gopalkrishna2000}, and double-double \citep{schoenmakers2000} sources have also been observed and there is still a continuing debate about the exact interplay between extrinsic effects, such as the interaction between the jet and the environment, and intrinsic effects, such as differences in central engines and accretion modes, that give rise to the different morphologies. In this work we use only the binary FRI/FRII classification.

We have used the MiraBest dataset 
which consists of 1256 images of radio galaxies pre-processed to be used specifically for deep learning tasks \citep[e.g.][]{bowles2021attention, e2cnn}. The dataset was constructed using the sample selection and classification described in \cite{Miraghaei2017TheEnvironment}, who made use of the parent galaxy sample from \cite{Best2012OnProperties}. Optical data from data release 7 of Sloan Digital Sky Survey \citep[SDSS DR7;][]{abazajian2009} was cross-matched with NRAO VLA Sky Survey  \citep[NVSS;][]{condon1998} and Faint Images of the Radio Sky at Twenty-Centimeters  \citep[FIRST;][]{becker1995} radio surveys. Parent galaxies were selected such that their radio counterparts had an active galactic nucleus (AGN) host rather than emission dominated by star formation. To enable classification of sources based on morphology, sources with multiple components in either of the radio catalogues were considered.

The morphological classification was done by visual inspection at three levels: (i) The sources were first classified as FRI/FRII based on the original classification scheme of \citet{fr1974}. Additionally, 35 \emph{Hybrid} sources were identified as sources having FRI-like morphology on one side and FRII-like on the other. Of the 1329 extended sources inspected, 40 were determined to be unclassifiable. (ii) Each source was then flagged as `Confident' or `Uncertain' to represent the degree of belief in the human classification. (iii) Some of the sources which did not fit exactly into the standard FRI/FRII dichotomy were given additional tags to identify their sub-type. These sub-types include 53 Wide Angle Tail (WAT), 9 Head Tail (HT) and 5 Double-Double (DD) sources. To represent these three levels of classification, each source was given a three digit identifier as shown in Table~\ref{tab:digits_mirabest}.

To construct the machine learning dataset, several pre-processing steps were applied to the data following the approach described in \citet{aniyan2017classifying} and \citet{tang2019}: 
\begin{enumerate}
    \item In order to minimise the background noise in the images, all pixels below the $ 3\,\sigma$ level of the background noise were set to 0. This threshold was chosen because among the classifiers trained by \citet{aniyan2017classifying} on images with $ 2\,\sigma$, $ 3\,\sigma$, and $ 5\,\sigma$ cut-offs, $ 3\,\sigma$ performed most well.
    \item The images were clipped to $150\times150$ pixels, centered on the source.
    \item The images were normalised as follows:
    \begin{equation}
        {\rm Output} = 255 ~ ~ \frac{ {\rm Input} - {\rm Input_{min}}} {{\rm Input_{max}} - {\rm Input_{min}}} ~ ,
        \label{eq:norm_mb}
    \end{equation}
    where ${\rm Input}$ refers to the input image, ${\rm Input_{min}}$ and ${\rm Input_{max}}$ are the minimum and maximum pixel values in the input image, and $ {\rm Output}$ is the image after normalisation.
\end{enumerate}

To ensure the integrity of the ML dataset, the following 73 objects out of the 1329 extended sources identified in the catalogue were not included: (i) 40 unclassifiable objects; (ii) 28 objects with extent greater than the chosen image size of $150\times150$ pixels; (iii) 4 objects which were found in overlapping regions of the FIRST survey; (iv) 1 object in category 103 (FRI Confident Diffuse). Since this was the only instance of this category, it would not have been possible for the test set to be representative of the training set. The composition of the final dataset is shown in Table~\ref{tab:mb_classes}. We do not include the sub-types in this table as we have not considered their classification.

In this work, we use the MiraBest Confident subset to train the BBB models. Examples of FRI and FRII galaxies from the MiraBest Confident dataset are shown in Figure~\ref{fig:Mirabest Confident}. 

Additionally, we use 49 samples from the MiraBest Uncertain subset and 30 samples from the MiraBest Hybrid class to test the trained model's ability to correctly represent different measures of uncertainty, since these samples can be considered as being drawn from the same data generating distribution as the MiraBest Confident samples, but have a differing degree of belief in their classification. We note that there may be components of either both epistemic and aleatoric uncertainty in the Uncertain and Hybrid samples, and this is discussed further in Section~\ref{sec:results}.




We note that all the previous work published using this dataset uses some form of data augmentation. In this work we do not use any data augmentation, although the effect of data augmentation is discussed further in Section~\ref{sec:augmentation}. The reasons for this are two-fold: firstly, that unprincipled data augmentation has been suggested to negatively affect the performance of Bayesian deep learning models \citep{nabarro2021data}; and secondly, that a noted advantage of Bayesian models is their ability to obtain good performance using only small datasets \citep[e.g.][]{xiong2011bayesian, semenova2020bayesian, jospin2020hands}.

\section{Model}
\label{sec:model}

\subsection{Architecture}
The architecture used to classify the MiraBest dataset using BBB is shown in Table~\ref{tab:mb_lenet_architecture}. We use a LeNet-5-style architecture \citep{lenet} with two additional convolutional layers. We found it essential to add two convolutional layers to the LeNet-5 architecture in order to obtain good model performance. Adding additional fully-connected and convolutional layers beyond this resulted in no further improvement. The number of channels in the additional convolutional layers were also optimised. We used a kernel support size of 5 to be consistent with previous CNN-style architectures used with the MiraBest dataset \citep[e.g.][]{e2cnn}. ReLU activation functions are used for each layer with the exception of the output layer and max-pooling is used to down-sample the feature data after each convolutional layer. 

The functional form of our priors is as defined in Section~\ref{sec:priors} and the hyper-parameters of the priors were tuned using the validation dataset. We build models with four different priors: (i) a simple Gaussian prior with $\sigma=0.1$, (ii) a Gaussian Mixture Model (GMM) prior with $\{\pi, \sigma_1, \sigma_2 \} = \{0.75, 1, ~ 9 . 10^{-4} \}$, (iii) a Laplace prior with $b = 1$, and (iv) a Laplace Mixture Model (LMM) prior with $\{\pi, b_1, b_2 \} = \{0.75, 1, 10^{-3} \}$.

We use a Gaussian distribution as our variational approximation to the posterior over both the weights and biases in our network. Models using the BBB method are known to be highly sensitive to the initialisation of this posterior and in this work we initialise the posterior means, $\mu$, from a uniform distribution, $\mathcal{U}(-0.1,0.1)$, and the posterior variance parameterisation, $\rho$, from $\mathcal{U}(-5,-4)$.

\begin{table}
\centering
\caption[CNN Architecture: MiraBest]{CNN architecture. Stride = 1 is used for all the convolutional and max pooling layers.}
	\begin{tabular}{lccc}
	\hline
    Operation & Kernel & Channels  & Padding\\
    \hline
    Convolution & 5 x 5 & 6 & 1 \\
    ReLU \\
    Max Pooling & 2 x 2 \\
    Convolution & 5 x 5 & 16 & 1 \\
    ReLU \\
    Max Pooling & 2 x 2 \\
    Convolution & 5 x 5 & 26 & 1 \\
    ReLU \\
    Max Pooling & 2 x 2 \\
    Convolution & 5 x 5 & 32 & 1 \\
    ReLU \\
    Max Pooling & 2 x 2 \\
    Fully-Connected & & 120 \\
    ReLU \\
    Fully-Connected & & 84 \\
    ReLU \\
    Fully-Connected & & 2 \\
    Log Softmax\\
    \hline
	\end{tabular}
   \label{tab:mb_lenet_architecture}
\end{table}

\subsection{Training}
\label{sec:training}

The MiraBest dataset has a predefined training:\,test split. We further divide the training data into a ratio of 80:20 to create training and validation sets. The final split contains 584 training samples, 145 validation samples and 104 test samples.

All the models are trained for 500 epochs, with minibatches of size 50. We train the models using the Adam optimiser with a learning rate of $\ell =  5 \times 10^{-5}$ for all the priors.
A learning rate scheduler is implemented which reduces the learning rate by 95\% if the validation likelihood cost does not improve for four consecutive epochs. 

After a model has been trained, the test error is calculated as the percentage of incorrectly classified galaxies by comparing the output of the model to the labels in the test set.

\subsection{Cold Posterior Effect}
\label{sec:cold}

It has been observed by several authors \citep{wenzel2020} that in order to get good predictive performance from Bayesian neural networks, the Bayesian posterior has to be down-weighted or tempered. We denote this by a weighting factor, $T$, in the cost function:
\begin{equation}
    F_i(D_i, \theta) = \frac{T}{M} {\rm KL} [q(w|\theta]||P(w)] - \mathbb{E}_{q(w|\theta)} [\log P(D_i|w)] ,
    \label{eq:tempered_cost_blundell}
\end{equation}
where $T \leq 1 $ is the temperature. We also observe this effect in our experiments.

We found it necessary to temper our posterior in order to get a good performance from the Bayesian neural network, without which the accuracy remains around $55\%$. We tempered the posterior for a range of temperature values, $T$, between $[10^{-5}, 1)$ and chose the largest value of $T$ for which the validation accuracy was improved significantly. Thus, for all experiments described in the following sections we use $T = 10^{-2}$ in Equation~\ref{eq:tempered_cost_blundell}.

Several hypotheses have been proposed to explain the cold posterior effect including model or prior misspecification \citep{wenzel2020}, and data augmentation or dataset curation leading to likelihood misspecification \citep{aitchison2021a}. These are discussed in detail and investigated further in Section~\ref{sec:masegosa}.

\subsection{Weight Pruning}
\label{sec:pruning}

Variational inference based neural networks have several advantages over non-Bayesian neural networks; however, for a typical variational posterior such as the Gaussian distribution, the number of parameters in the network double compared to a non-Bayesian model with the same architecture because both the mean and standard deviation values need to be learned. This increases the computational and memory overhead at test time and during deployment. Thus, there is a need to develop network pruning approaches which can be used to remove the parameters that contain least or no useful information. Several authors have also considered pruning to improve the generalisation performance of the network \citep{lecun1990optimal}. Many of the pruning methods that have been developed can also be applied to non-Bayesian neural networks, but in this section we discuss a Signal-to-Noise ratio (SNR) based pruning criterion which can be applied naturally to a model trained with Gaussian variational densities  \citep{practicalvi, blundell}. 

The SNR of a model weight is calculated as follows:
\begin{equation}
    {\rm SNR} = \frac{|\mu|} { \sigma} , 
    \label{eq:snr}
\end{equation}
where $\mu$ and $\sigma$ are the values of the variational parameters after a model has been trained. In practice, we use the SNR values in dB:
\begin{equation}
    {\rm SNR}_{{\rm dB}} = 10 \log_{10} {\rm SNR} . 
    \label{eq:snr_db}
\end{equation}
%
The weights of the network with the lowest SNR values are removed. The effect of removing these parameters is measured by re-calculating the test error using the pruned model. While the SNR-based method may seem very simple, it allows for a large proportion of weights to be removed for some models. We discuss alternative pruning approaches in Section~\ref{sec:pruning2}.

We adapt SNR-based pruning for a convolutional Bayesian neural network by considering only the fully-connected layer weights of the model for pruning, instead of all the weights of the network. This is because the convolutional layer weights are shared weights and removing even a small fraction may result in disastrous consequences for model performance. However, the fully-connected layers make up $\sim 85\%$ of the total weights of our network, so pruning methods are still worth considering for convolutional BBB models. Pruning only the fully-connected layers is also consistent with pruning methods developed for standard CNN models  \citep[e.g.][]{gong2014compressing,soulie2016compression, tu2016reducing}.

For the model trained on the MiraBest dataset with a Laplace prior, we find that up to 30\% of the fully-connected layer weights can be pruned without a significant change in performance. This is discussed further in Section~\ref{sec:pruning2}.

\section{Results}
\label{sec:results}

\subsection{Classification and Calibration}

\begin{table}
\centering
\caption[Classification error on MiraBest using BBB-CNN]{Classification error and percentage classwise Uncertainty Calibration Error (cUCE) on MiraBest Confident test set using BBB-CNN. The percentage cUCE is shown separately for predictive uncertainty as measured by predictive entropy (PE), epistemic uncertainty as measured by mutual information (MI) and aleatoric uncertainty as measured by average entropy (AE) as calculated on the MiraBest Confident test set. For a fuller explanation of these metrics, please see Section~\ref{sec:uncertainty}.}
	\begin{tabular}{lcccc}
	 &  & \multicolumn{3}{c}{\textbf{cUCE \%}} \\\cmidrule{3-5}
    \textbf{Prior}  & \textbf{Test Error} &  \textbf{PE} & \textbf{MI}& \textbf{AE}  \\
    \hline
    Gaussian Prior  & $14.48 \pm 3.40 \%$ &  30.49 & 21.90 & 25.48 \\
    GMM Prior  & $ 12.89 \pm 2.23 \%$ & 19.92 & 18.86 & 16.86 \\
    Laplace Prior  & $ 11.62 \pm 2.38 \%$ & 9.69 & 16.37 & 10.84 \\
    LMM Prior  & $ 17.29 \pm  2.71 \%$ & 21.02 & 26.05 & 17.69\\
	\end{tabular}
   \label{tab:results_mirabest_bbb}
\end{table}

\begin{figure*}
    \centering
    \includegraphics[width=\textwidth]{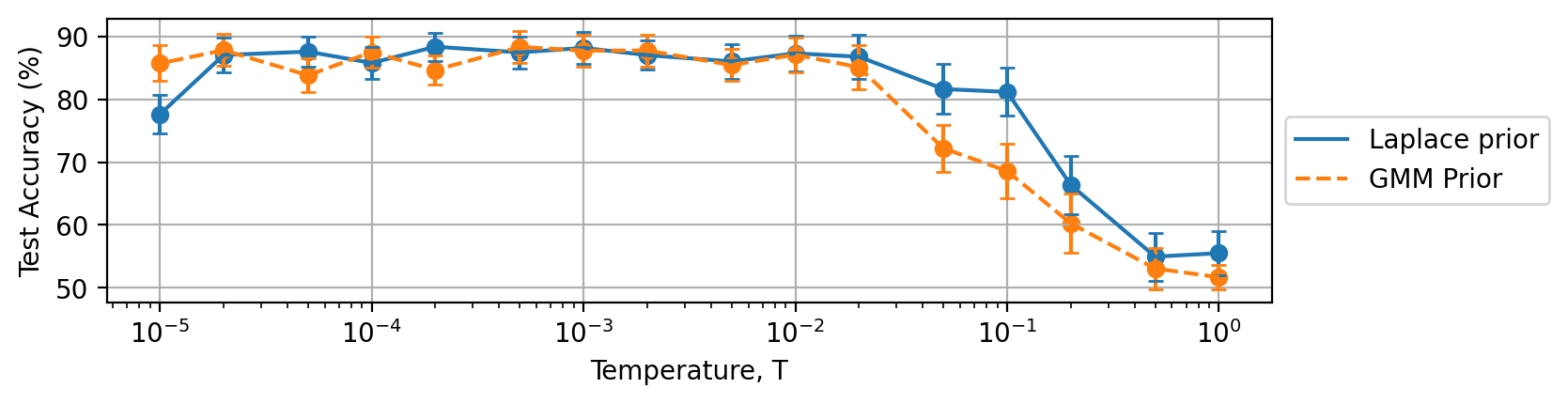}
    \caption{The “cold posterior” effect for the MiraBest classification problem, see Section~\ref{sec:masegosa} for details. Data are shown for the BBB models with no data augmentation and the original ELBO cost function trained with a Laplace prior (solid blue line), and trained with a GMM prior (orange dashed line).}
    \label{fig:prior_cold}
\end{figure*}

The results of our classification experiments are shown in Table~\ref{tab:results_mirabest_bbb}. The mean and standard deviation values of the test error are calculated by taking 100 samples from the posterior predictive distribution for each test data point in the MiraBest Confident data set.

Using the model trained with a Laplace prior, we recover a test error of $11.62 \pm 2.38 \%$. We also recover a comparable test error of $12.89 \pm 2.23 \%$ using a GMM prior. The standard deviation values represent the spread in the overall test error and indicate the model's confidence in its predictions on the test set. \citet{bowles2021attention} who augment the MiraBest Confident samples by a factor of 72 report a test error of $8\%$, whereas \citet{e2cnn} who use random rotations of the same dataset as a function of epoch to augment the data, report a test error of $5.95 \pm 1.37 \%$ with a LeNet-5 style CNN with Monte Carlo dropout, and $3.43\pm 1.29 \%$ using a $D_{16}$ group-equivariant CNN with Monte Carlo dropout. 
We again emphasise that the test error values we report are without any data augmentation. 
If we include data augmentation using random rotations from 0 to 360\,degrees this improves the BBB test error using the Laplace prior to $7.41 \pm 2.22\%$, but at the cost of increased uncertainty calibration error. We note that the wider effect of using augmentation with Bayesian models is a subject of debate in the literature and this is discussed further in Section~\ref{sec:augmentation}.

Whilst differences in performance are often used to choose a preferred model, it is also the case that more accurate point-wise models are \emph{over-confident} in their predictions. This problem of over-confidence in standard NNs is well documented in the literature, see e.g. \cite{nguyen2014}. In particular this effect has been shown to lead to miscalibrated uncertainty in predictions, especially for data samples that are less similar to canonical examples of a class \citep{guo2017,hein2019}. 



Table~\ref{tab:results_mirabest_bbb} shows the percentage classwise Uncertainty Calibration Error (cUCE) values of the uncertainty metrics calculated for the MiraBest Confident test set, see Section~\ref{sec:calibration}. Among the four priors tested in this work, we find that the Laplace prior gives the most well-calibrated uncertainty metrics, followed by the GMM prior. Given the uncertainties on each test error due to the small size of our test set, we cannot draw any strong conclusions about which prior should be preferred. However, after analysing the uncertainty calibration error for each prior, we suggest that the Laplace prior produces the most well-calibrated uncertainties. We also find that the cold posterior effect is less pronounced in the case of the Laplace prior model, see Figure~\ref{fig:prior_cold}. Similarly to the model accuracy, these results indicate that learning benefits most from a sparser prior and consequently in the following analysis we report results for the Laplace prior.

\subsection{Uncertainty Quantification}

\begin{figure*}
\begin{center}
    \begin{subfigure}[b]{0.23\textwidth}
        \includegraphics[width=\textwidth]{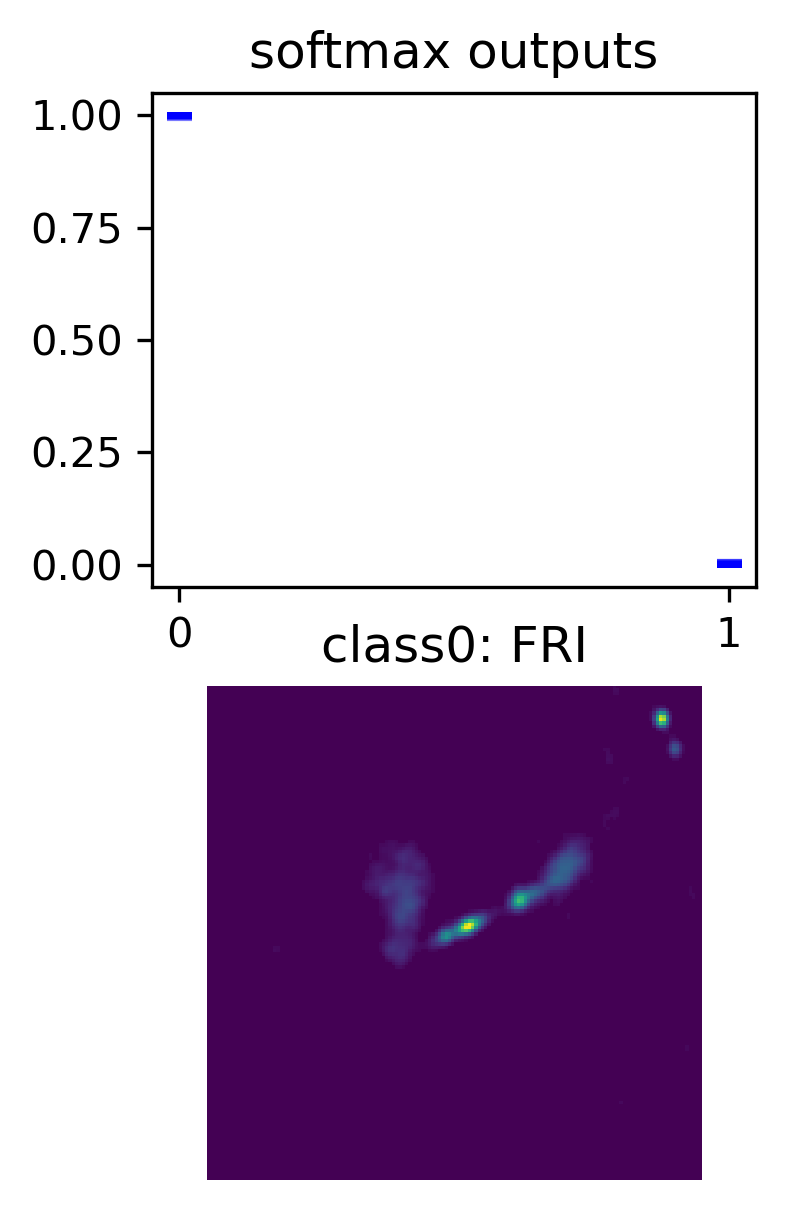}
        \caption[]{}
        \label{fig:A_correct}
    \end{subfigure} %
	\hfill
	\begin{subfigure}[b]{0.23\textwidth}
	    \includegraphics[width=\textwidth]{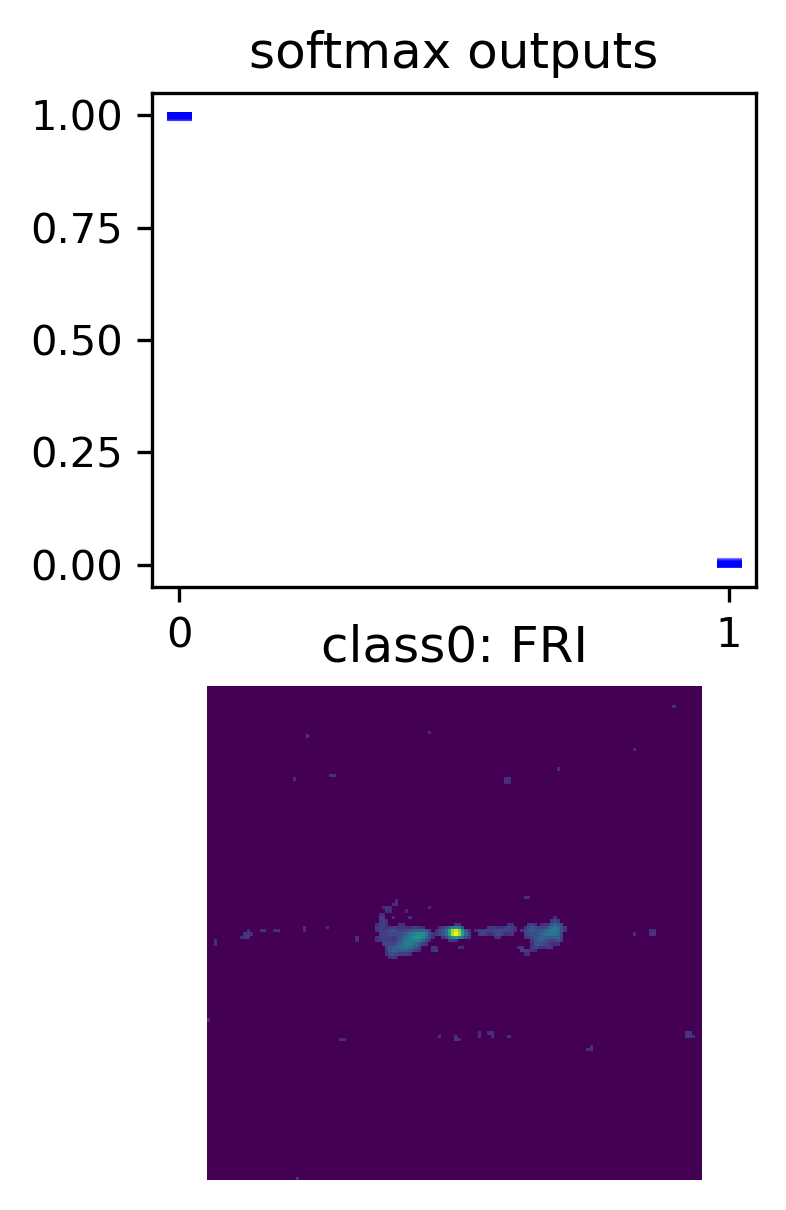}
        \caption[]{}
        \label{fig:B_correct}
	\end{subfigure}
	\hfill
	\begin{subfigure}[b]{0.23\textwidth}
	    \includegraphics[width=\textwidth]{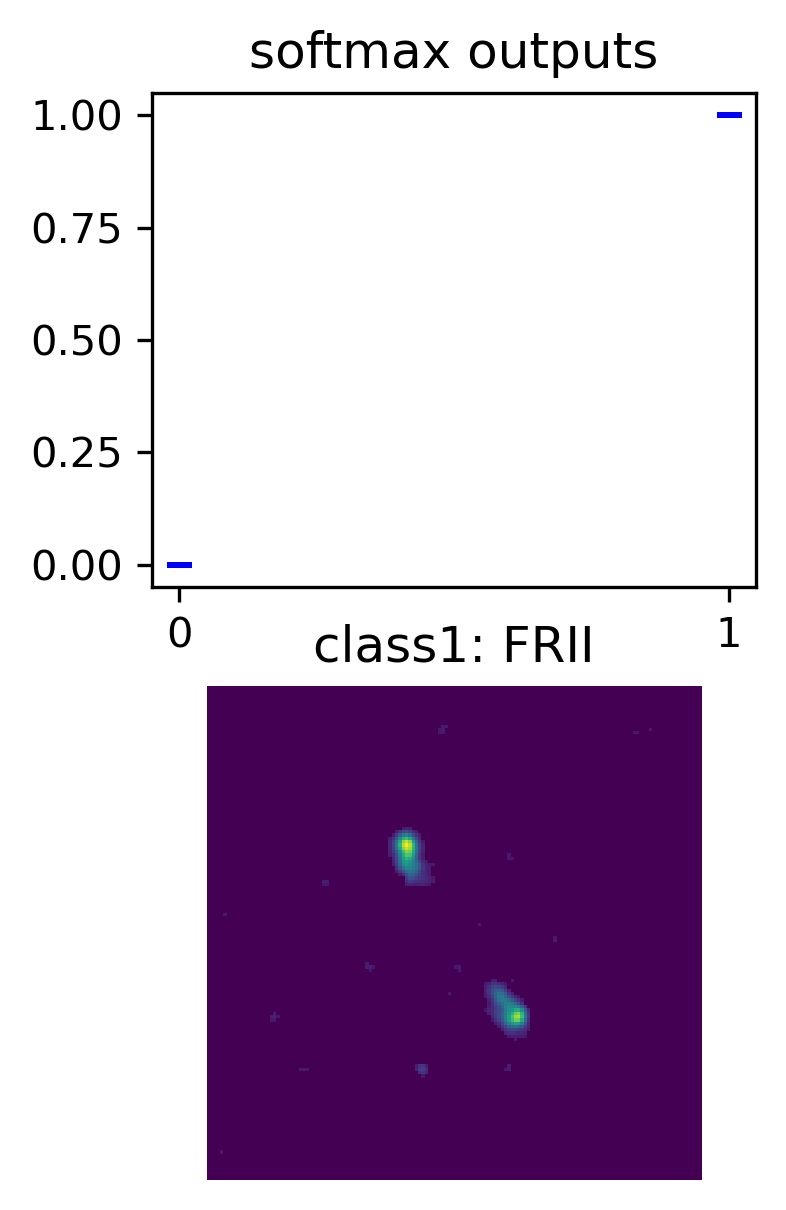}
        \caption[]{}
        \label{fig:C_correct}
	\end{subfigure}
	\hfill
	\begin{subfigure}[b]{0.23\textwidth}
	    \includegraphics[width=\textwidth]{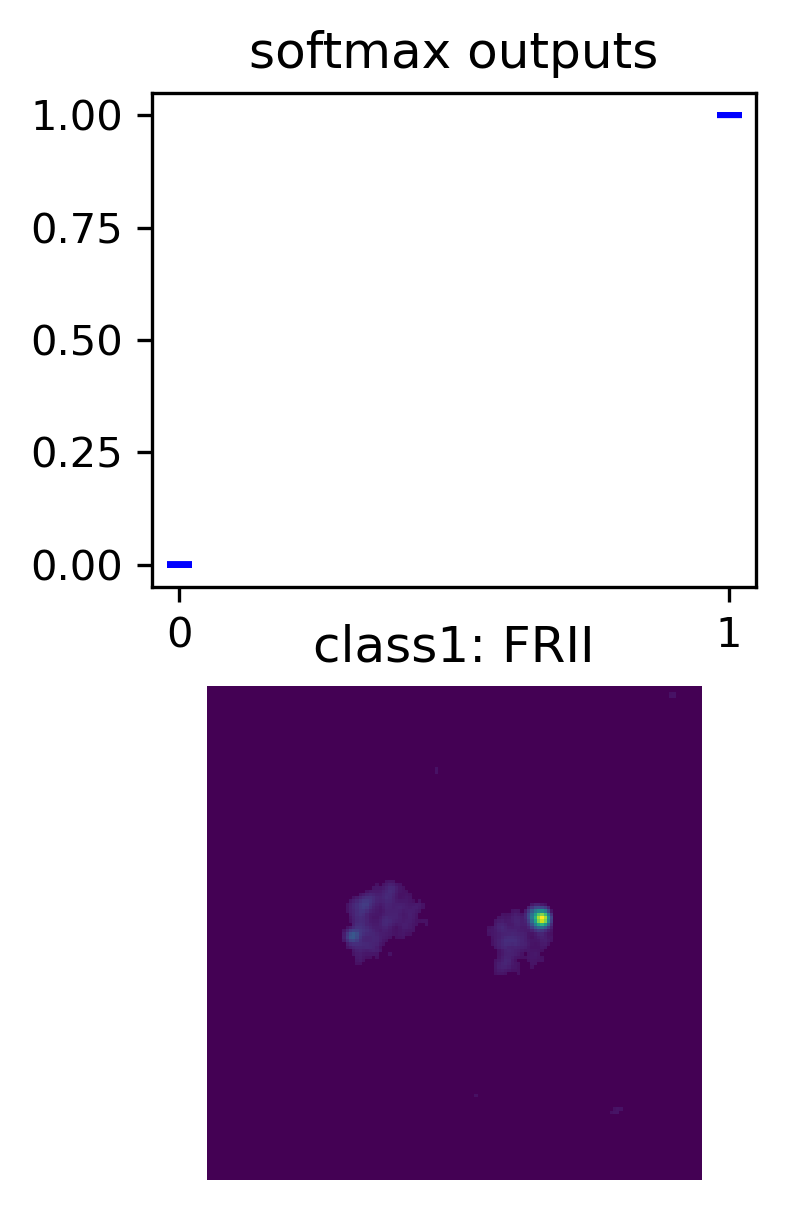}
        \caption[]{}
        \label{fig:D_correct}
	\end{subfigure}
	\caption{Examples of galaxies correctly classified with high predictive confidence. Top: softmax values for 200 forward passes through the trained model. Bottom: input data images.}
	\label{fig:correct_highconf} 
\end{center}
\end{figure*}

\begin{figure*}
\begin{center}
    \begin{subfigure}[b]{0.23\textwidth}
        \includegraphics[width=\textwidth]{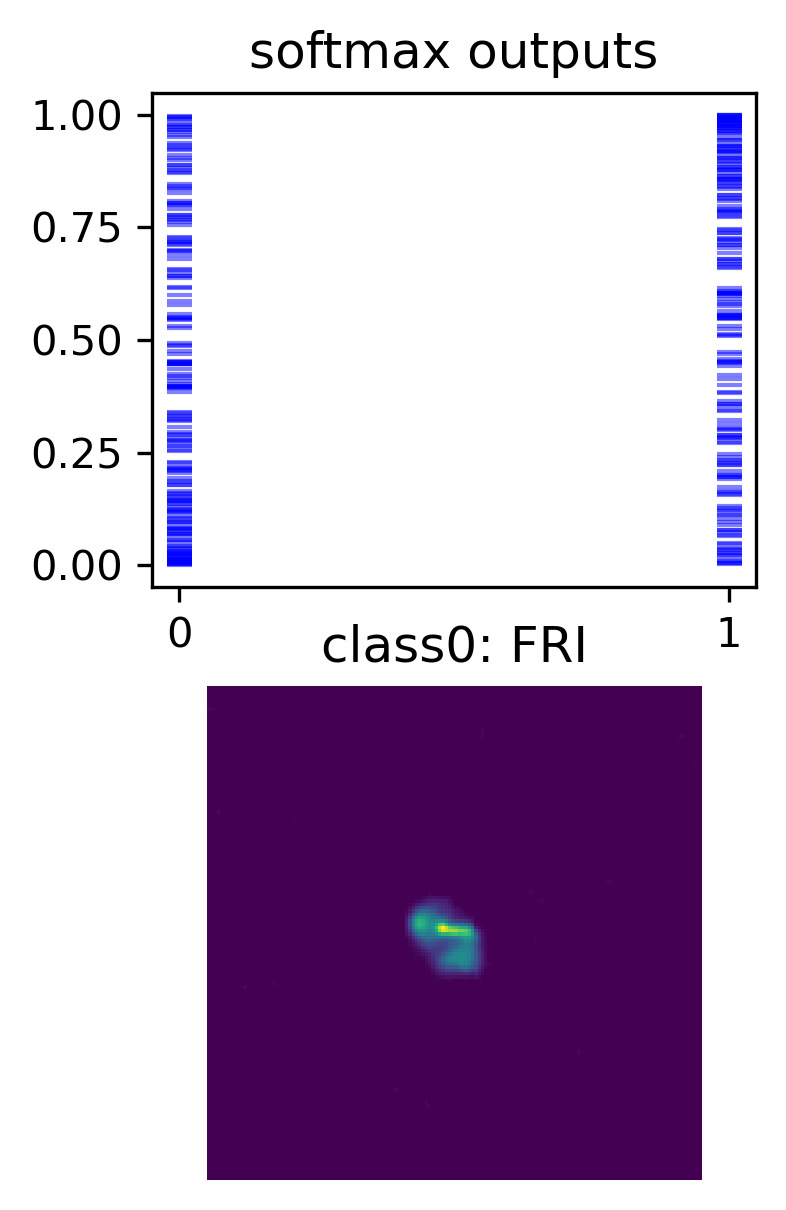}
        \caption[]{}
        \label{fig:A_highent}
    \end{subfigure} %
	\hfill
	\begin{subfigure}[b]{0.23\textwidth}
	    \includegraphics[width=\textwidth]{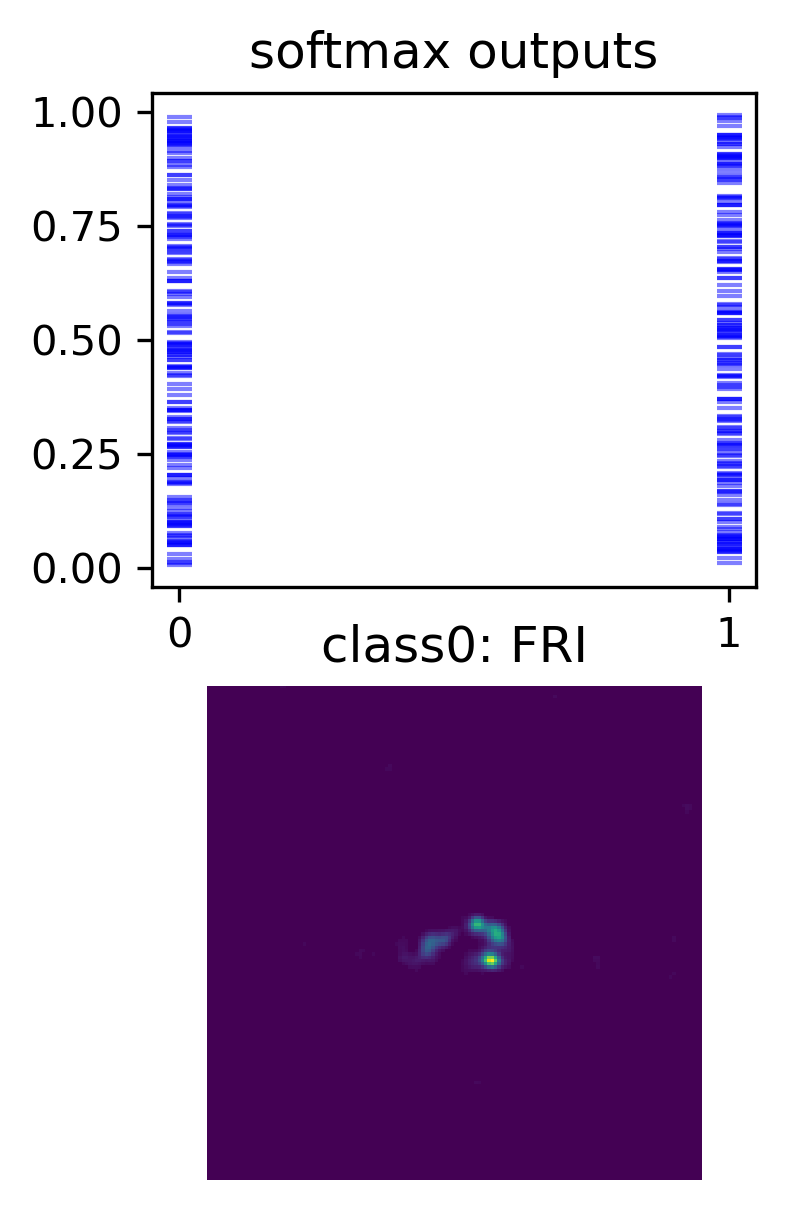}
        \caption[]{}
        \label{fig:B_highent}
	\end{subfigure}
	\hfill
	\begin{subfigure}[b]{0.23\textwidth}
	    \includegraphics[width=\textwidth]{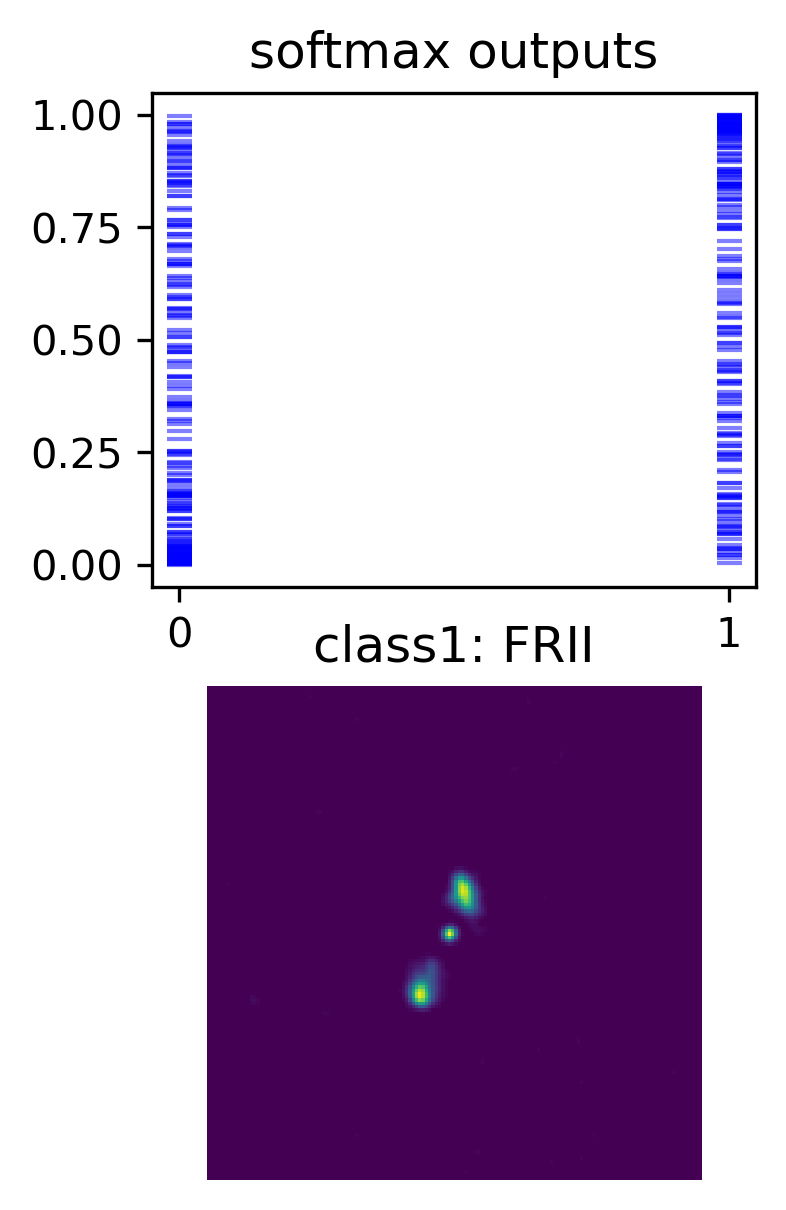}
        \caption[]{}
        \label{fig:C_highent}
	\end{subfigure}
	\hfill
	\begin{subfigure}[b]{0.23\textwidth}
	    \includegraphics[width=\textwidth]{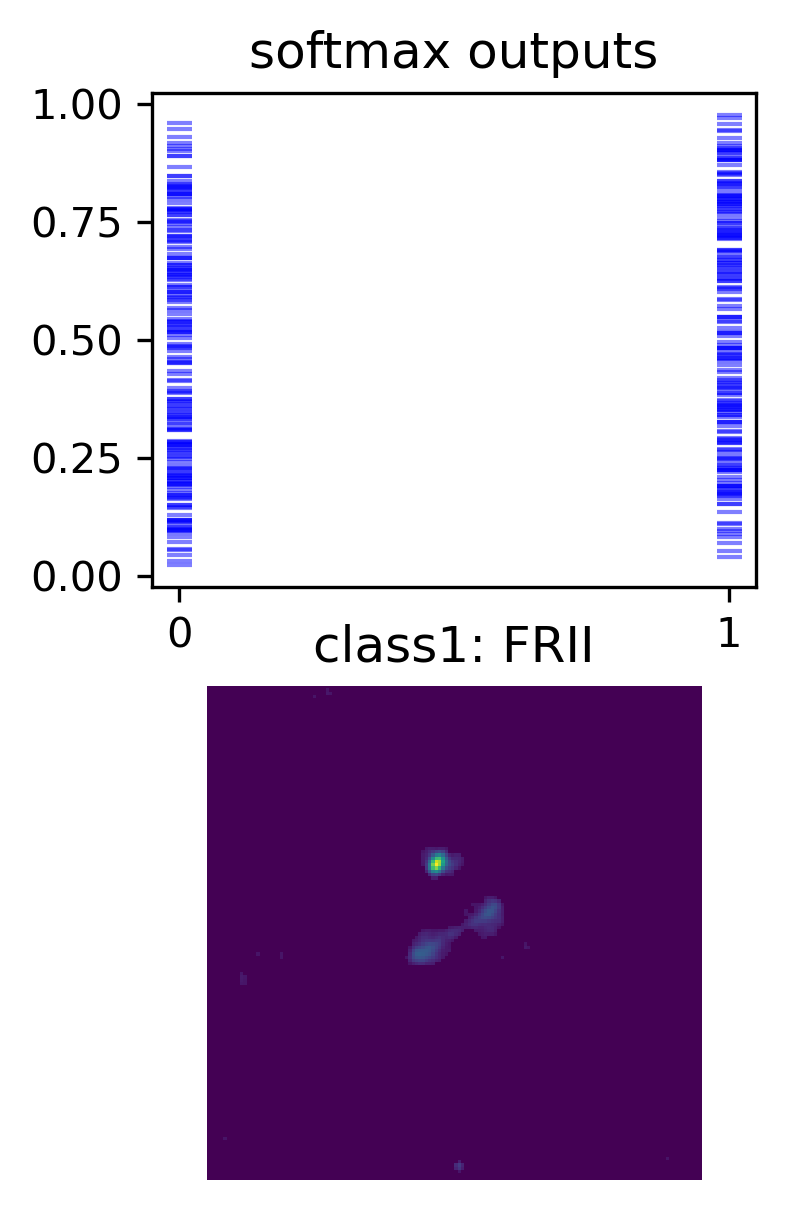}
        \caption[]{}
        \label{fig:D_highent}
	\end{subfigure}
	\caption{Examples of galaxies classified with low predictive confidence. Top: softmax values for 200 forward passes through the trained model. Bottom: input data images.}
	\label{fig:highent} 
\end{center}
\end{figure*}

\begin{figure}[]
    \centering
    \includegraphics[width=0.23\textwidth]{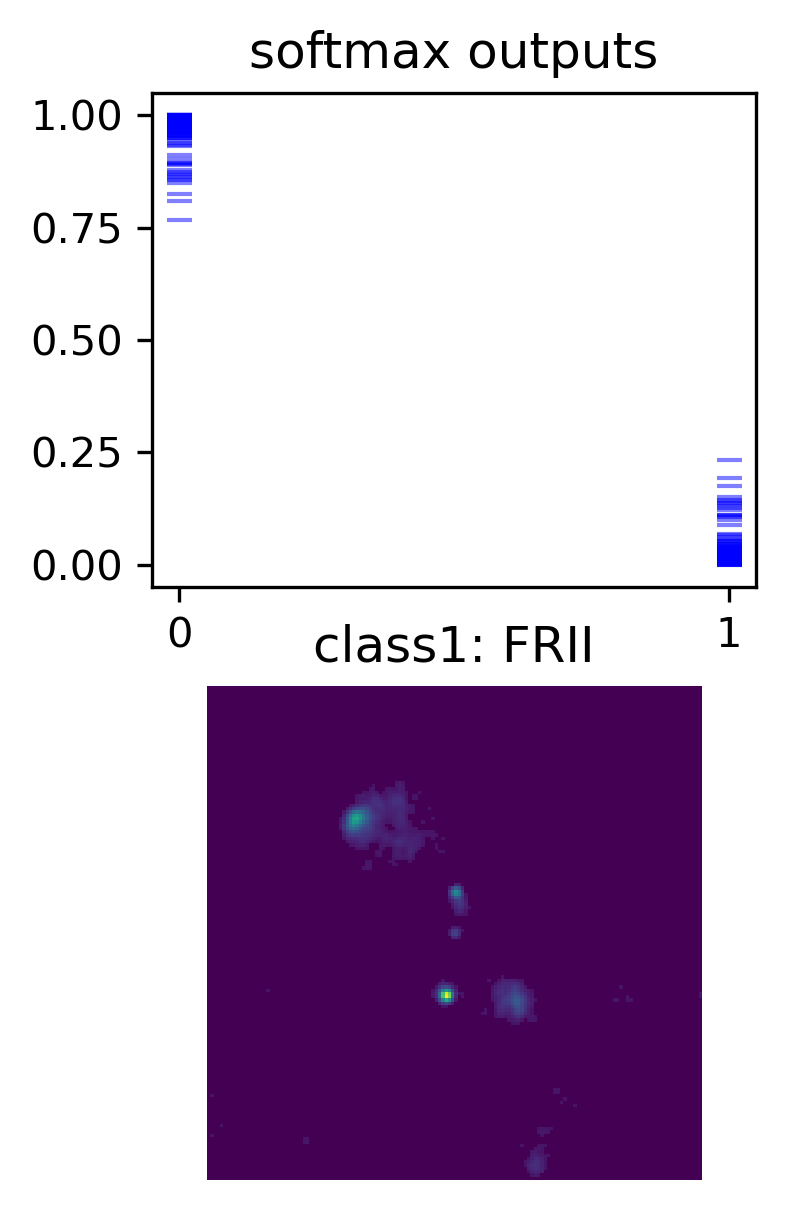}
    \caption{A galaxy that has been incorrectly classified with high predictive confidence. Top: softmax values for 200 forward passes through the trained model. Bottom: input data image.}
    \label{fig:incorrect_highconf}
\end{figure}


We first look at some illustrative examples of galaxies from the MiraBest Confident test set to build intuition about the uncertainty metrics used. For each test sample we make $N = 200$ forward passes through the trained model. This results in a distribution of model outputs following the learned predictive posterior, which allow us to estimate the different uncertainty measures described in Section~\ref{sec:uncertainty}.

We show some examples of galaxies that have been correctly classified with high confidence in Figure~\ref{fig:correct_highconf}. These galaxies correspond to typical FRI/FRII classifications. The corresponding uncertainty metrics are shown in Table~\ref{tab:low_ent}. The predictive entropy and mutual information for all the galaxies shown are very low ($< 0.01$ nats). The overlap indices $\eta_{\rm soft}$ and $\eta_{\rm logits}$ are $\ll 10^{-5}$, which indicates that there is virtually no overlap, as can be seen in the distributions of Softmax probabilities in Figure~\ref{fig:correct_highconf}.

We then consider galaxies for which the predictive uncertainty is high, as shown in Figure~\ref{fig:highent}. These galaxies have the highest predictive entropy among the test samples of the MiraBest Confident dataset and large values of overlap indices in both the softmax and logit space. These samples also have high mutual information, which indicates that the model's confidence in its classification is very low. The values of uncertainty metrics corresponding to these galaxies are shown in Table~\ref{tab:high_ent}.

Finally in Figure~\ref{fig:incorrect_highconf}, we show one example where the model has incorrectly classified a galaxy with high confidence. The galaxy has been labelled an FRII and the model incorrectly classifies it as an FRI. The predictive entropy, mutual information and overlap indices are very low, as shown in Table~\ref{tab:incorrect_highconf}, which means that the model's confidence in its prediction is very high for this galaxy. We can see that the galaxy deviates from the typical FRII classification because it has additional bright components and its label is somewhat ambiguous. Thus, the bias introduced by the ambiguity in the definition of FRI/FRII and the ambiguity in the labels gives rise to uncertainty metrics that can potentially be misleading. 

In this section we saw how high or low values of predictive entropy, mutual information and overlap indices indicate the model's confidence in making predictions about individual galaxies; in the next section we analyse the distributions of uncertainty metrics for all the galaxies in the dataset.


\begin{table}
\centering
\caption[Uncertainties in samples correctly classified with high confidence]{Predictive entropy (PE), mutual information (MI) and overlap indices for Softmax ($\eta_{\rm soft}$) and logit-space ($\eta_{\rm logits}$) for galaxies correctly classified with high confidence shown in Figure~\ref{fig:correct_highconf}}
	\begin{tabular}{ccccc}
    \textbf{Galaxy} & \textbf{PE} & \textbf{MI} & $\boldsymbol{\eta_{\rm soft}}$ & $\boldsymbol{\eta_{\rm logits}}$ \\
    \hline
    3a & $<$0.01 & $<$0.01 & $\ll 10^{-5}$ & $\ll 10^{-5}$ \\
    3b & $<$0.01 & $<$0.01 & $\ll 10^{-5}$ & $\ll 10^{-5}$ \\
    3c & $<$0.01 & $<$0.01 & $\ll 10^{-5}$ & $\ll 10^{-5}$ \\
    3d & $<$0.01 & $<$0.01 & $\ll 10^{-5}$ & $\ll 10^{-5}$ \\
    \end{tabular}
  \label{tab:low_ent}
\end{table}

\begin{table}
\centering
\caption[Uncertainties in samples incorrectly classified with low confidence]{Predictive entropy (PE), mutual information (MI) and overlap indices for Softmax ($\eta_{\rm soft}$) and logit-space ($\eta_{\rm logits}$) for galaxies classified with low confidence shown in Figure~\ref{fig:highent}.}
	\begin{tabular}{ccccc}
    \textbf{Galaxy} & \textbf{PE} & \textbf{MI} & $\boldsymbol{\eta_{\rm soft}}$ & $\boldsymbol{\eta_{\rm logits}}$ \\
    \hline
    4a & 0.68 & 0.25 & 0.72 & 0.10 \\
    4b & 0.69 & 0.20 & 0.90 & 0.08 \\
    4c & 0.67 & 0.27 & 0.70 & 0.08 \\
    4d & 0.69 & 0.14 & 0.88 & 0.12 \\
    \end{tabular}
  \label{tab:high_ent}
\end{table}

\begin{table}
\centering
\caption[Uncertainties in samples incorrectly classified with high confidence]{Predictive entropy (PE), mutual information (MI) and overlap indices for Softmax ($\eta_{\rm soft}$) and logit-space ($\eta_{\rm logits}$) for a galaxy incorrectly classified with high confidence shown in Figure~\ref{fig:incorrect_highconf}.}
	\begin{tabular}{ccccc}
    \textbf{Galaxy} & \textbf{PE} & \textbf{MI}  & $\boldsymbol{\eta_{\rm soft}}$ & $\boldsymbol{\eta_{\rm logits}}$\\
    \hline
    5 & 0.10 & 0.02 & $<$0.01 & $<$0.01 \\
    \end{tabular}
   \label{tab:incorrect_highconf}
\end{table}

\subsection{Analysis of Uncertainty Estimates}

\begin{figure*}
\begin{center}
    \begin{subfigure}[b]{0.29\textwidth}
        \includegraphics[width=\textwidth]{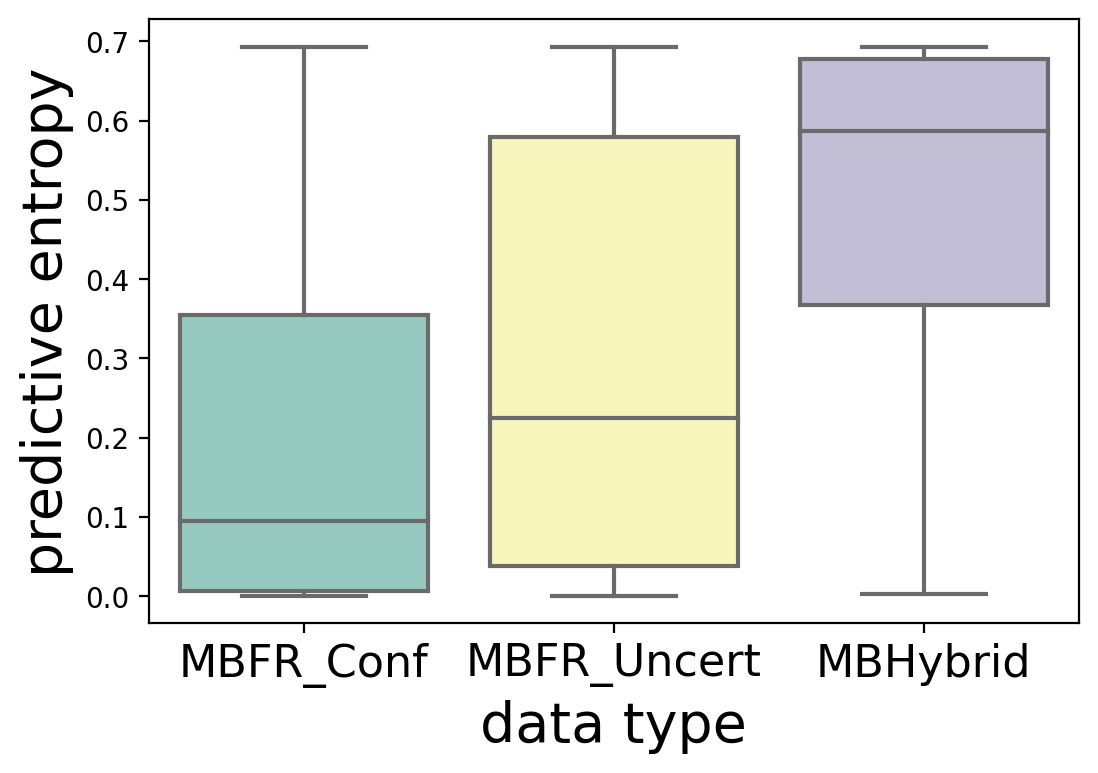}
        \caption[]{}
        \label{fig:all_entropy}
    \end{subfigure} %
	\hfill
	\begin{subfigure}[b]{0.29\textwidth}
	    \includegraphics[width=\textwidth]{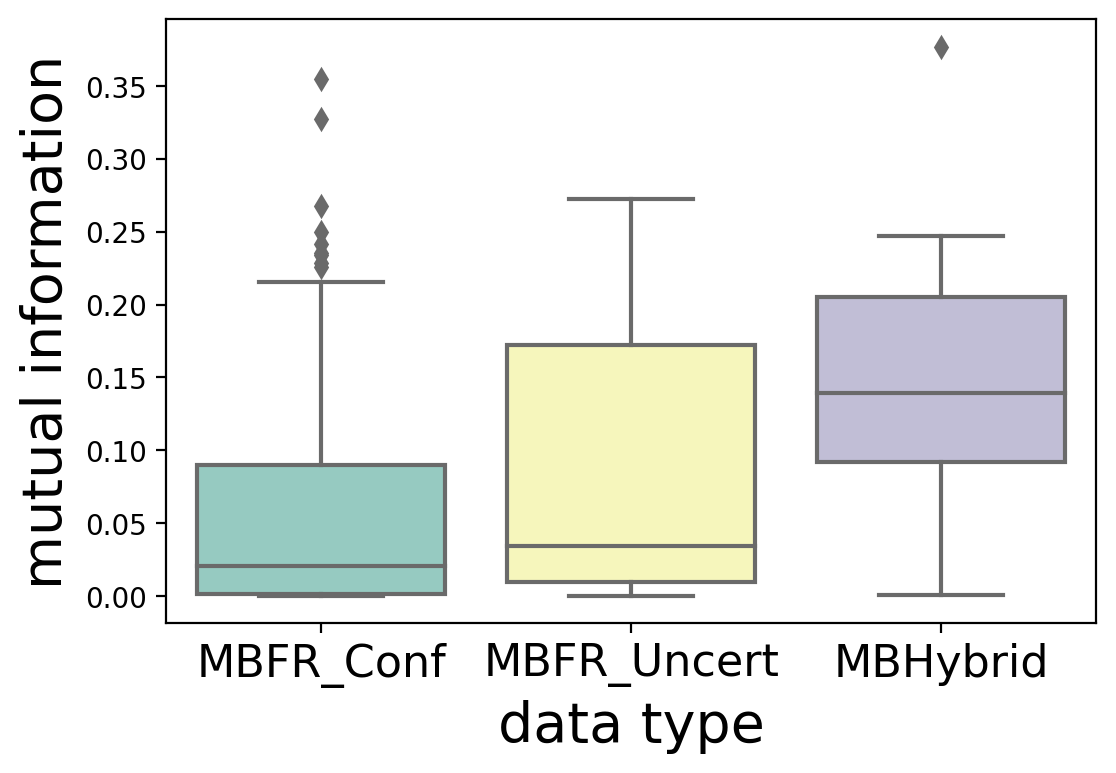}
        \caption[]{}
        \label{fig:all_MI}
	\end{subfigure}
	\hfill
	\begin{subfigure}[b]{0.29\textwidth}
	    \includegraphics[width=\textwidth]{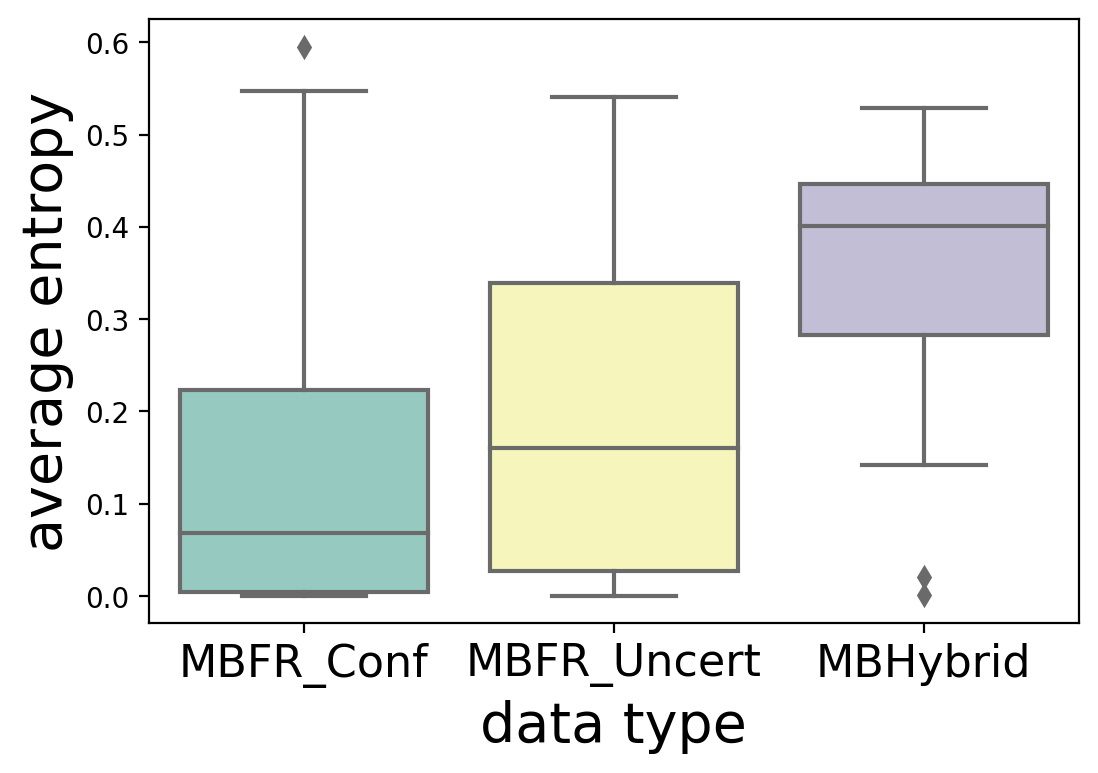}
        \caption[]{}
        \label{fig:all_aleat}
	\end{subfigure}
	\caption{Distributions of uncertainty metrics for MiraBest Confident (MBFR\_Conf), Uncertain (MBFR\_Uncert) and Hybrid (MBHybrid) datasets. (a) Predictive uncertainty as measured using predictive entropy; (b) Epistemic uncertainty as measured using mutual information; and (c) Aleatoric uncertainty as measured using average entropy. For a fuller explanation of these metrics, please see Section~\ref{sec:uncertainty}.}
	\label{fig:all_uncert} 
\end{center}
\end{figure*}

\begin{figure*}
\begin{center}
    \begin{subfigure}[b]{0.33\textwidth}
        \includegraphics[width=\textwidth]{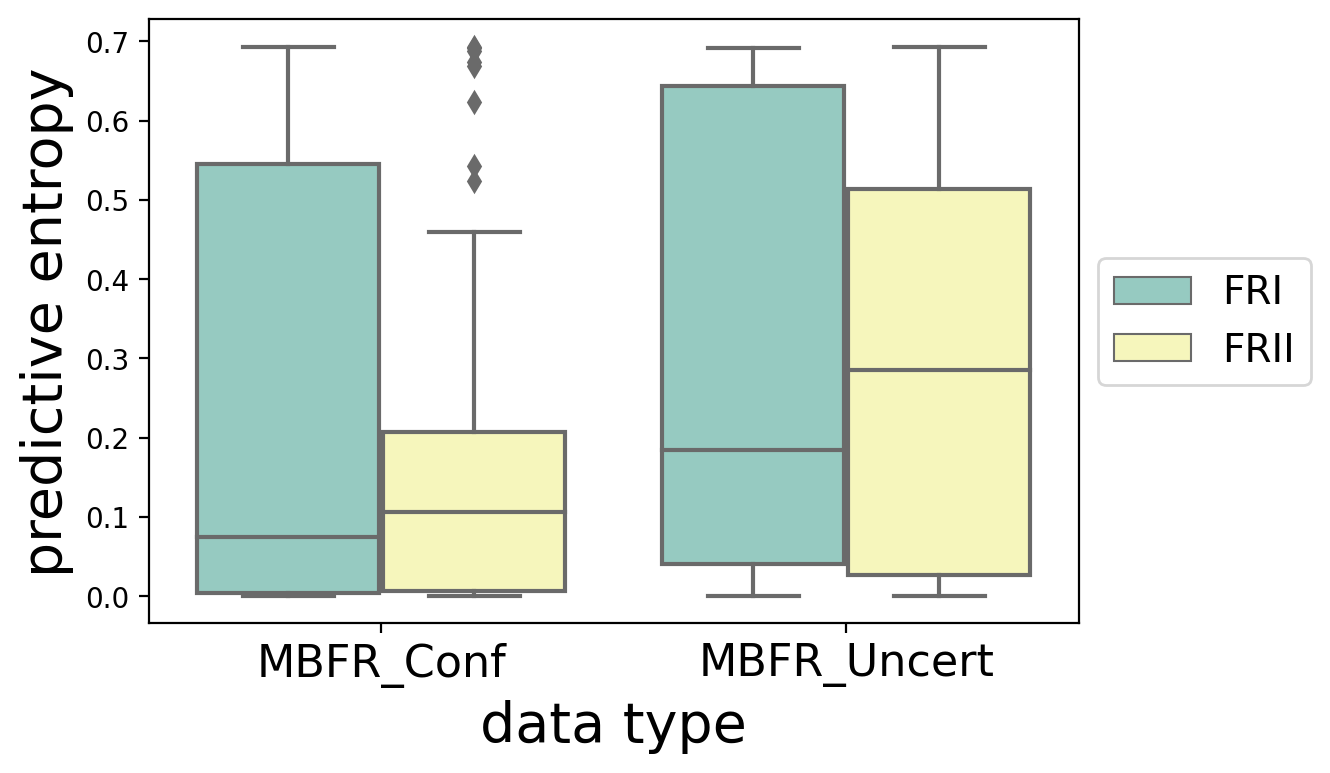}
        \caption[]{}
        \label{fig:conf_vs_uncert_entropy}
    \end{subfigure} %
	\hfill
	\begin{subfigure}[b]{0.33\textwidth}
	    \includegraphics[width=\textwidth]{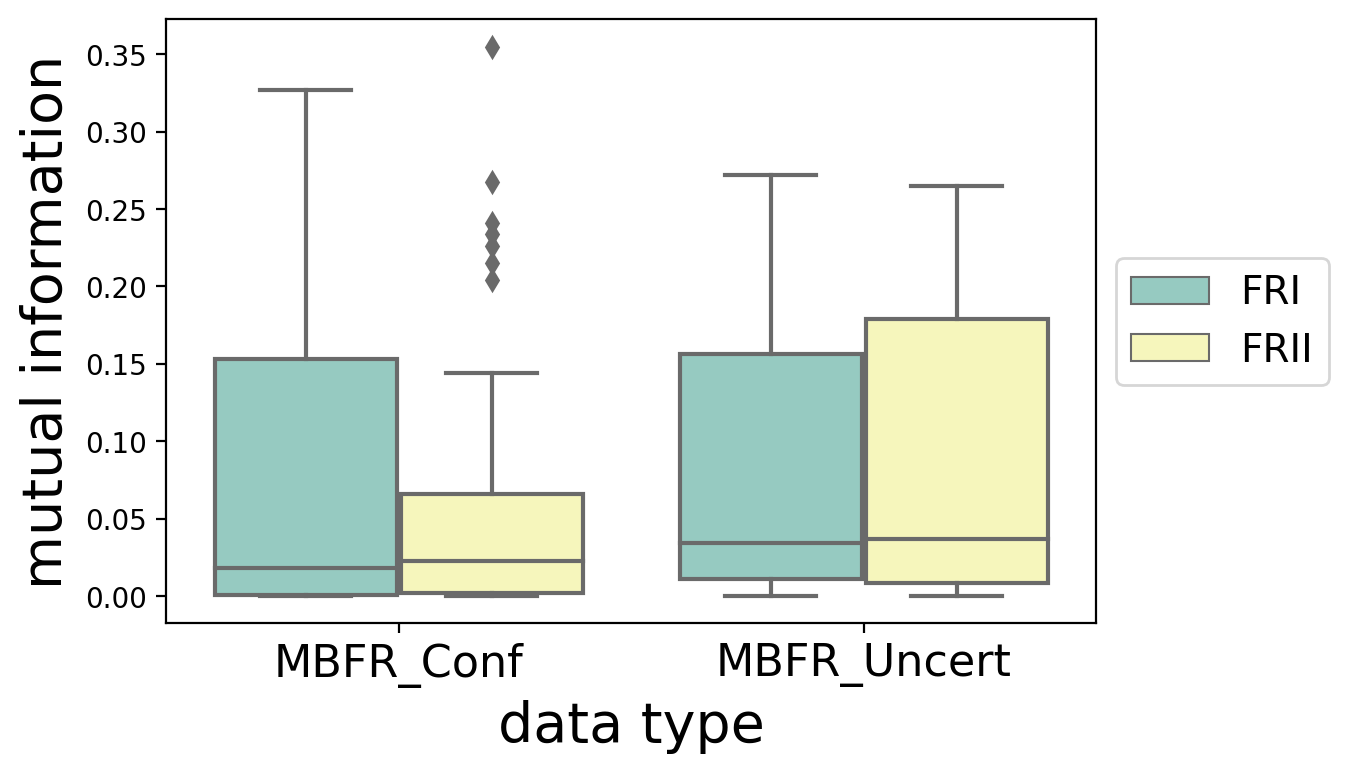}
        \caption[]{}
        \label{fig:conf_vs_uncert_MI}
	\end{subfigure}
	\hfill
	\begin{subfigure}[b]{0.33\textwidth}
	    \includegraphics[width=\textwidth]{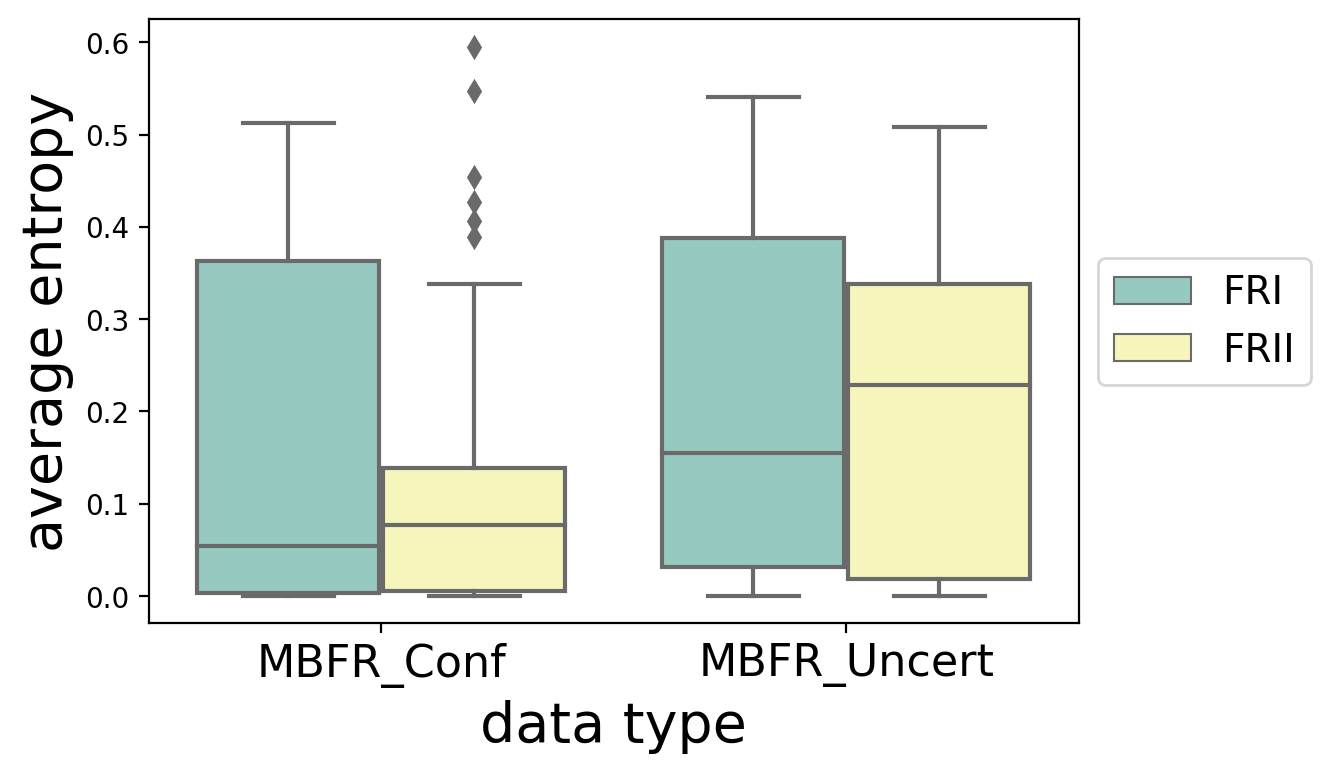}
        \caption[]{}
        \label{fig:conf_vs_uncert_aleat}
	\end{subfigure}
	\caption{Class-wise distributions of uncertainty metrics for MiraBest Confident and MiraBest Uncertain datasets. (a) Predictive uncertainty as measured using predictive entropy; (b) Epistemic uncertainty as measured using mutual information; and (c) Aleatoric uncertainty as measured using average entropy. For a fuller explanation of these metrics, please see Section~\ref{sec:uncertainty}.}
	\label{fig:conf_vs_uncert} 
\end{center}
\end{figure*}

\begin{figure*}
\begin{center}
    \begin{subfigure}[b]{0.3\textwidth}
        \includegraphics[width=\textwidth]{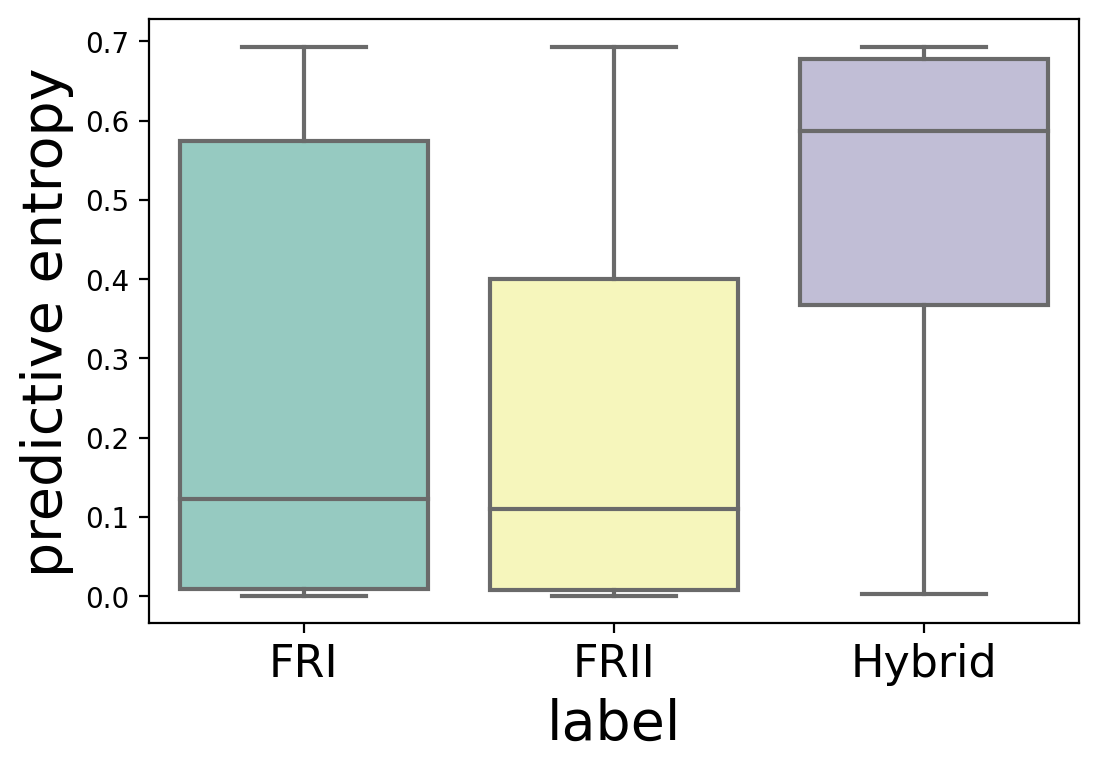}
        \caption[]{}
        \label{fig:morph_entropy}
    \end{subfigure} %
	\hfill
	\begin{subfigure}[b]{0.3\textwidth}
	    \includegraphics[width=\textwidth]{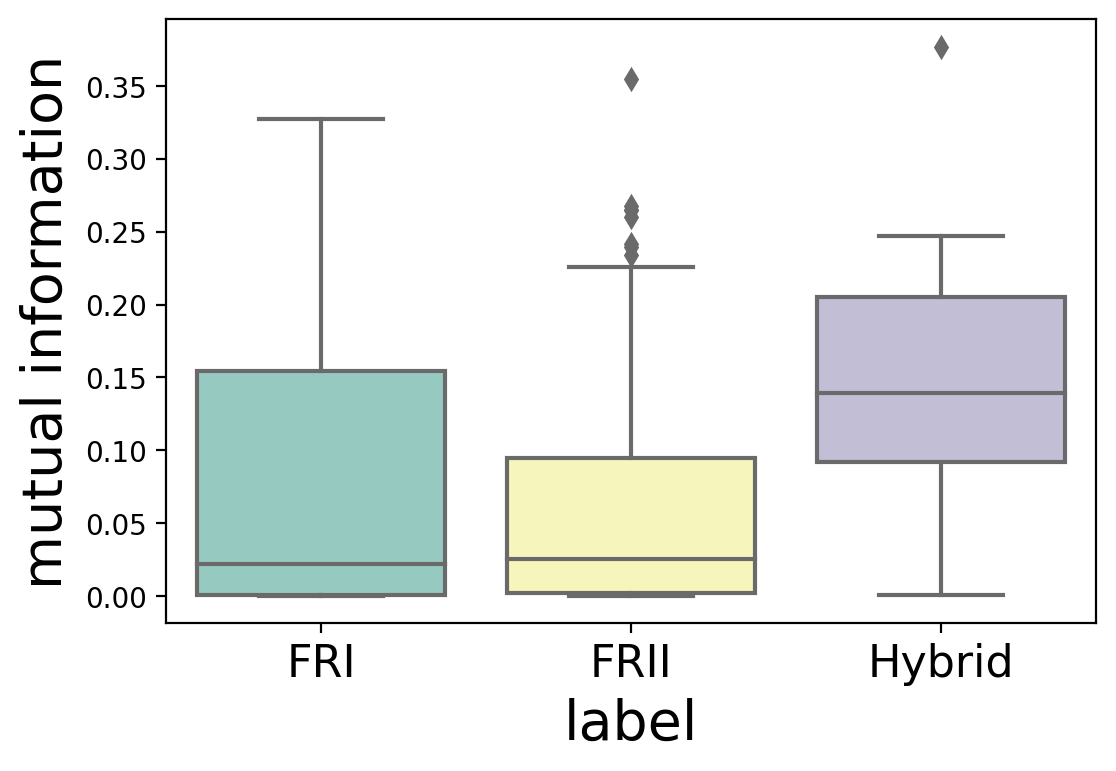}
        \caption[]{}
        \label{fig:morph_MI}
	\end{subfigure}
	\hfill
	\begin{subfigure}[b]{0.3\textwidth}
	    \includegraphics[width=\textwidth]{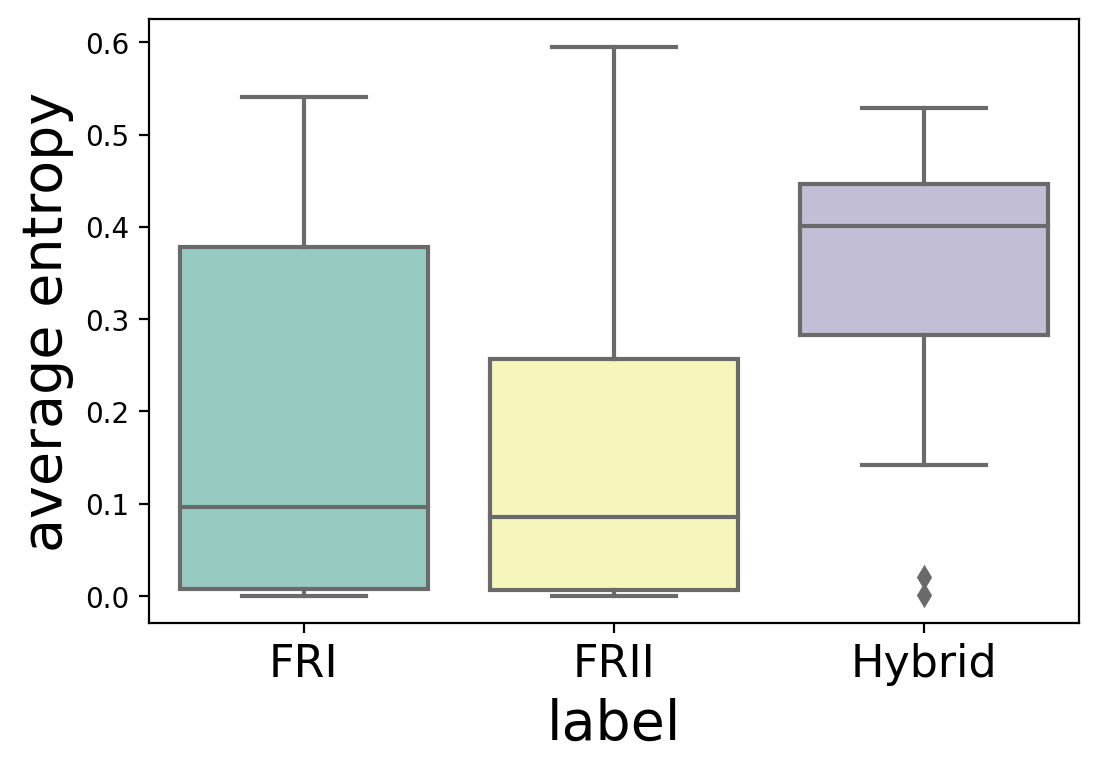}
        \caption[]{}
        \label{fig:morph_aleat}
	\end{subfigure}
	\caption{Morphology-wise distributions of uncertainty metrics for the MiraBest dataset. (a) Predictive uncertainty as measured using predictive entropy; (b) Epistemic uncertainty as measured using mutual information; and (c) Aleatoric uncertainty as measured using average entropy. For a fuller explanation of these metrics, please see Section~\ref{sec:uncertainty}.}
	\label{fig:morph_uncert} 
\end{center}
\end{figure*}

\begin{figure*}
\begin{center}
    \begin{subfigure}[b]{0.33\textwidth}
        \includegraphics[width=\textwidth]{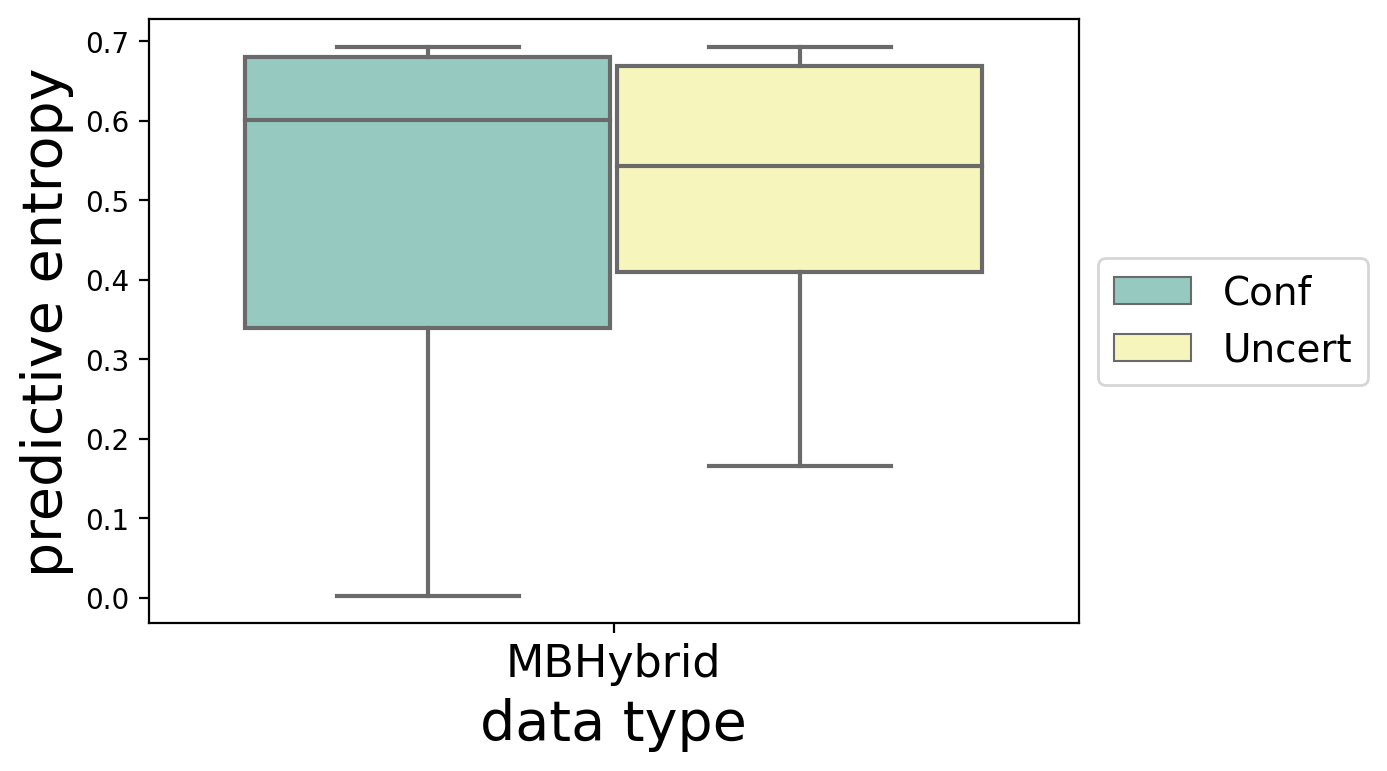}
        \caption[]{}
        \label{fig:hybrid_uncert_entropy}
    \end{subfigure} %
	\hfill
	\begin{subfigure}[b]{0.33\textwidth}
	    \includegraphics[width=\textwidth]{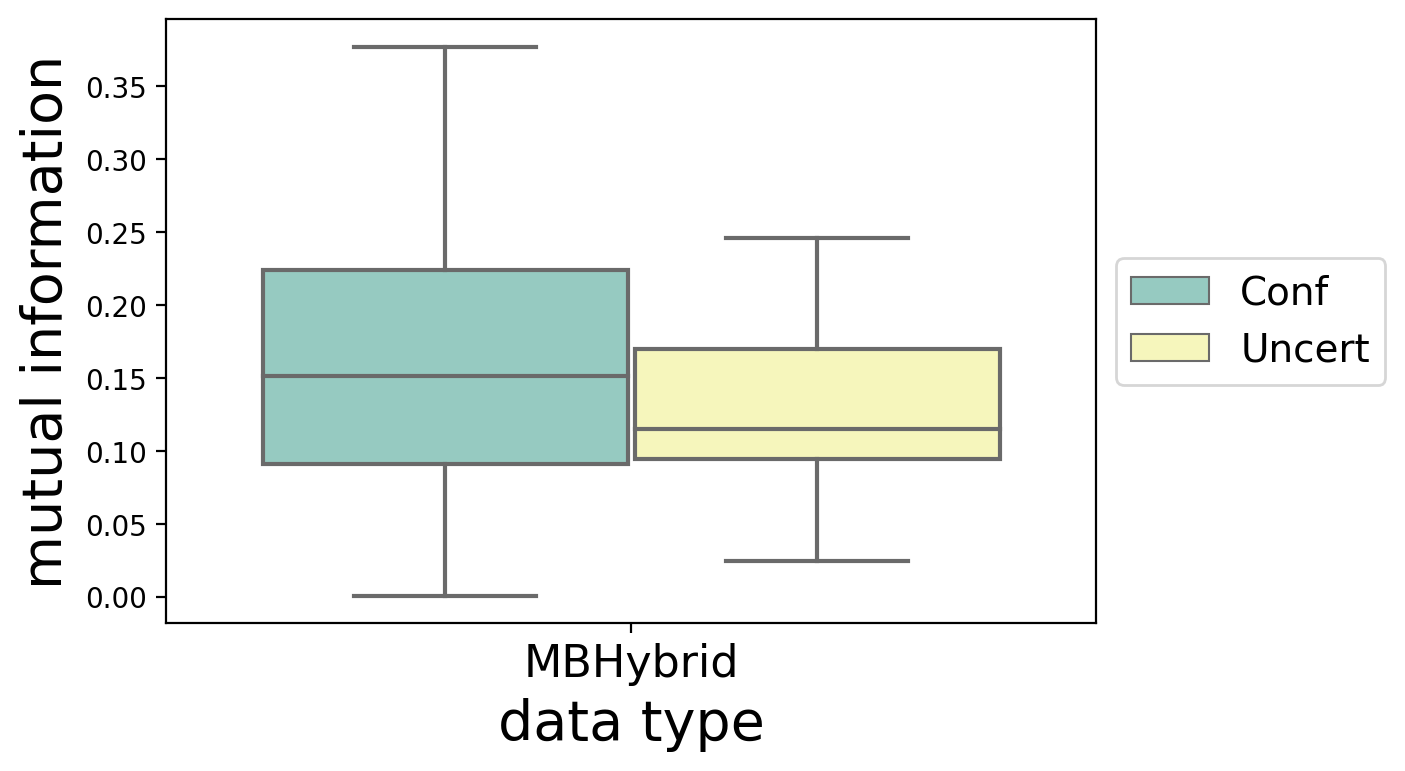}
        \caption[]{}
        \label{fig:hybrid_uncert_MI}
	\end{subfigure}
	\hfill
	\begin{subfigure}[b]{0.33\textwidth}
	    \includegraphics[width=\textwidth]{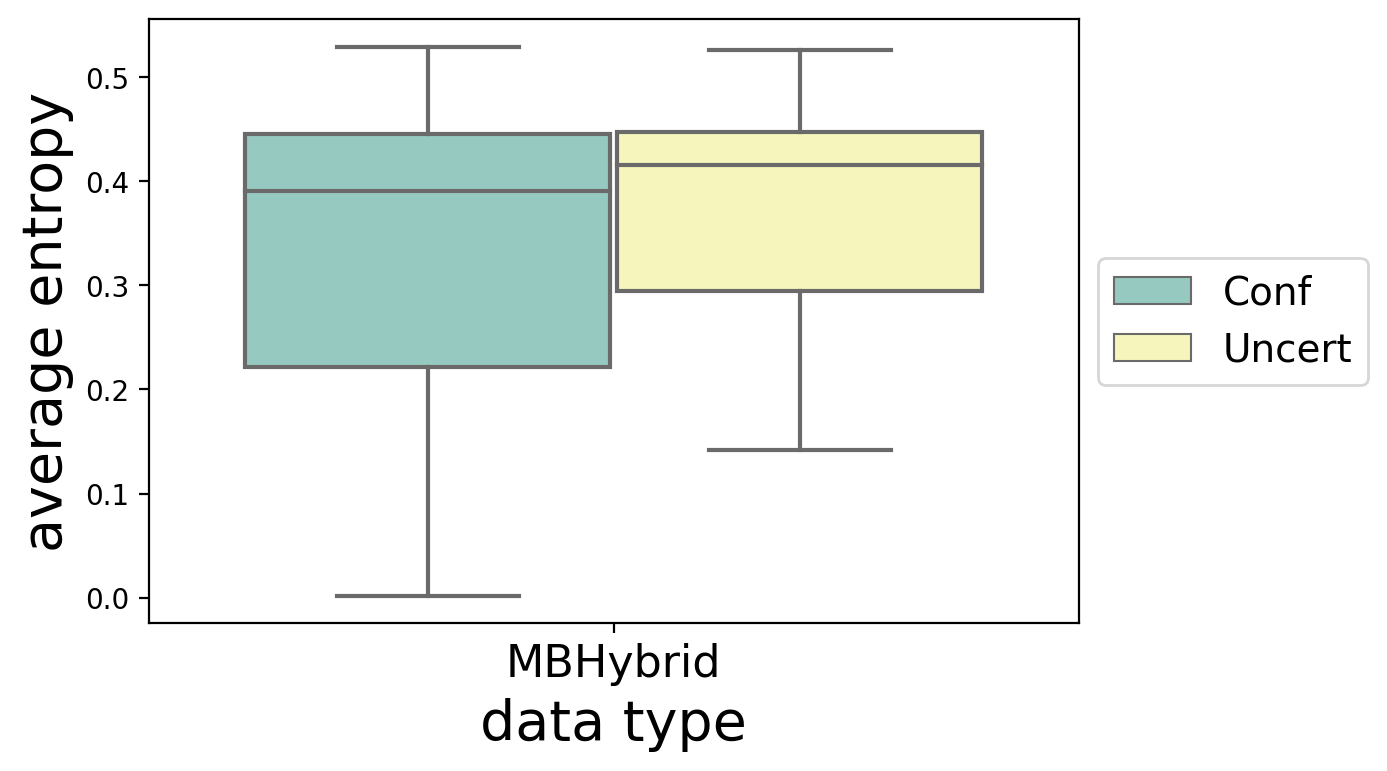}
        \caption[]{}
        \label{fig:hybrid_uncert_aleat}
	\end{subfigure}
	\caption{Class-wise distributions of uncertainty metrics for MiraBest Hybrid dataset. (a) Predictive uncertainty as measured using predictive entropy; (b) Epistemic uncertainty as measured using mutual information; and (c) Aleatoric uncertainty as measured using average entropy. For a fuller explanation of these metrics, please see Section~\ref{sec:uncertainty}.}
	\label{fig:hybrid_uncert} 
\end{center}
\end{figure*}

We test the trained model's ability to capture different measures of uncertainty by calculating uncertainty metrics for (i) the MiraBest Uncertain test samples (49 objects), and (ii) the MiraBest Hybrid samples (30 objects), using the model trained on the MiraBest Confident samples. As before this is done by making $N = 200$ forward passes through the model for each test input. Overall distributions for each uncertainty metric as a function of the three different test sets are shown in Figure~\ref{fig:all_uncert}.

The MiraBest Uncertain samples are considered to be drawn from the same distribution as the MiraBest Confident samples, and from a machine learning perspective would therefore be denoted as \emph{in- distribution} (iD). The MiraBest Hybrid samples are a more complex case: in principle they are a separate class that was not considered when training the model and therefore might be denoted as being \emph{out-of-distribution} (OoD) by some measures; however, given that they are still a sub-population of the over-all radio galaxy population, and moreover that they are defined as amalgams of the two classes used to train the model, they could also be considered to be in-distribution. Consequently, in this work we treat the MiraBest Hybrid test sample as being in-distribution.

\subsubsection{Analysis of Uncertainty Estimates on MiraBest Uncertain}
\label{sec:mbuncert}

Figure~\ref{fig:all_uncert} shows that the MiraBest Uncertain test set has on average higher measures of uncertainty across all estimators than the MiraBest Confident test set. In Figure~\ref{fig:conf_vs_uncert_entropy} we can see that this is also reflected in the median values of the predictive entropy distribution being higher for both FRI and FRII classes in the MiraBest Uncertain test set compared to the Confident test set, and that the interquartile range is also larger. This indicates that a larger number of galaxies are being classified with higher predictive entropy. The predictive entropy distribution has a higher median value for FRII objects for both MiraBest Confident and Uncertain samples. However, the distributions are wider for FRI objects.


The distribution of mutual information is shown in Figure~\ref{fig:all_MI}. The mutual information is also higher for samples from the MiraBest Uncertain test set. This indicates a higher epistemic uncertainty in classifying these samples, which is consistent with how the datasets are defined. We also find that FRIIs have a wider distribution of epistemic uncertainty than FRIs for MiraBest Uncertain samples, see Figure~\ref{fig:conf_vs_uncert_MI}. 

Using the average entropy of a test sample as a measure of aleatoric uncertainty, we see in
Figure~\ref{fig:all_aleat} that the Uncertain samples have higher median value and a larger interquartile range than the Confident samples. From Figure~\ref{fig:conf_vs_uncert_aleat} we note that that for both FRI and FRII type galaxies, the interquartile range has shifted to a higher value and the median average entropy is also higher.


From Figure~\ref{fig:morph_uncert}, we can also see that FRII samples have lower uncertainty than FRI samples when combined across the Confident and Uncertain test sets, which could be because the training set contains $\sim7$\% more FRIIs than FRIs.

Thus in general we find that BBB can correctly represent \emph{model} uncertainty in radio galaxy classification, and that this uncertainty is correlated with how human classifiers defined the MiraBest Confident and Uncertain qualifications.

\subsubsection{Analysis of Uncertainty Estimates on MiraBest Hybrid}
\label{sec:mbhybrid}

In Figure~\ref{fig:all_uncert} we can see that the interquartile range of the distributions of uncertainty metrics for the MiraBest Hybrid samples are well-separated from the distributions for the MiraBest Confident samples.

The median predictive entropy of the Hybrid samples is higher than the MiraBest Confident samples by $0.5$ nats, as shown in Figure~\ref{fig:all_entropy}. This indicates that there is a high degree of predictive uncertainty associated with the hybrid samples, which is expected as the training set does not contain any hybrid samples. This behaviour is also echoed as a function of overall morphology, see Figure~\ref{fig:morph_uncert}. It can be seen that the MiraBest Hybrid samples have substantially higher median uncertainties than either the FRI or FRII objects (combined across the MiraBest Confident and MiraBest Uncertain samples).

In Figure~\ref{fig:all_MI} we see that the median value for the distribution of mutual information for the Hybrid test set is higher than the upper quartile of the MiraBest Confident test set. This high degree of epistemic uncertainty could be because the model did not see any Hybrid samples during training.
We also note that among the sub-classes of the Hybrid dataset, the confidently labelled samples have higher epistemic uncertainty than the uncertainly labelled samples, as shown in Figure~\ref{fig:hybrid_uncert_MI}. We suggest that this may be because the uncertain samples are more similar to the FRI/FRII galaxies that the model has seen during training, i.e. their classification as a Hybrid was considered uncertain by a human classifier because the morphology was biased towards one of the standard FRI or FRII classifications. In which case their epistemic uncertainty might be expected to be lower since the model was trained to predict those morphologies.


The distribution of the average entropy (aleatoric uncertainty) has a higher median value and the interquartile range has shifted to higher values of average entropy compared to the Confident and Uncertain test sets, see Figure~\ref{fig:all_aleat}. While it can be seen that Hybrid samples have higher aleatoric uncertainty on average, in Figure~\ref{fig:hybrid_uncert_aleat} we can also see how the aleatoric uncertainty is distributed among the classes in the hybrid samples. The confidently labelled Hybrid samples span almost the entire range of the entropy function between $(0, 0.693]$ nats. The uncertainty labelled samples also have a high degree of aleatoric uncertainty.

Thus we find that the Hybrid test set has an even higher degree of uncertainty than the Uncertain test set.

\subsection{Alternative pruning approaches}
\label{sec:pruning2}

In Section~\ref{sec:pruning} we found that 30\% of the weights in the fully-connected layers of our trained model could be pruned without a loss of performance using an SNR-based pruning approach. In this section, we discuss an alternative pruning approach based on Fisher information. We compare the performance of our BBB model trained on radio galaxies for different pruning methods and analyse the effect of pruning on uncertainty estimates.

A number of alternative methods for model pruning have been described in the literature. Hessian based methods have been proposed to rank model parameters by their importance, but in practice the calculation of a full Hessian for typical deep learning models is in general prohibitively expensive in terms of computation. Consequently, one of the most popular methods is to simply rank parameters by their magnitude, where magnitude refers to the absolute value of the weights. The SNR approach, where parameters with low magnitudes or high variances are removed (Section~\ref{sec:pruning}), may be considered a natural extension of this method. \cite{tu2016reducing} showed that for deterministic neural networks it is possible to improve on a simple magnitude-based pruning by using the Fisher information matrix (FIM) for a particular parameter of the network, $\theta$. Here we implement the method of \cite{tu2016reducing} and compare it to the SNR-based pruning method.

These two approaches are based on fundamentally different methodologies: while SNR pruning takes into account "noisy" weights that are either too small in magnitude or have large posterior variances, the Fisher-information based method removes parameters based on their contribution to the gradients. If a parameter has smaller FIM values, this indicates that the gradients of the parameter did not change much during training, i.e. that the parameter contained less information and was less relevant to producing the optimised model. Thus one method may be preferred to the other in specific applications.

The empirical FIM for a parameter, $\theta$, can be calculated as follows:
\begin{equation}
    F(\theta) = \mathbb{E}_y \left[ \left( \frac{\partial \log L}{\partial \theta}\right) \left( \frac{\partial \log L}{\partial \theta}\right)^T \right] ~ , 
\end{equation}
where $L$ is the loss function. \cite{tu2016reducing} used the log likelihood loss function for their classical neural network, but for our model this loss is the ELBO function, see Equation~\ref{eq:elbo}. Using the Adam optimiser to train the models as described in Section~\ref{sec:training} allows one to use the bias-corrected second raw moment estimate of the gradient to approximate the FIM diagonal, since this value is used by the Adam optimiser to adapt to the geometry of the data \citep{adam}. The value of the FIM diagonal for each parameter is used to rank the parameters in order of importance and those with the lowest values are removed. For a more thorough discussion of different FIM calculations used in the statistical and machine learning literature we refer the reader to \cite{kunstner}. 

\begin{figure*}
\begin{center}
	\includegraphics[width=\textwidth]{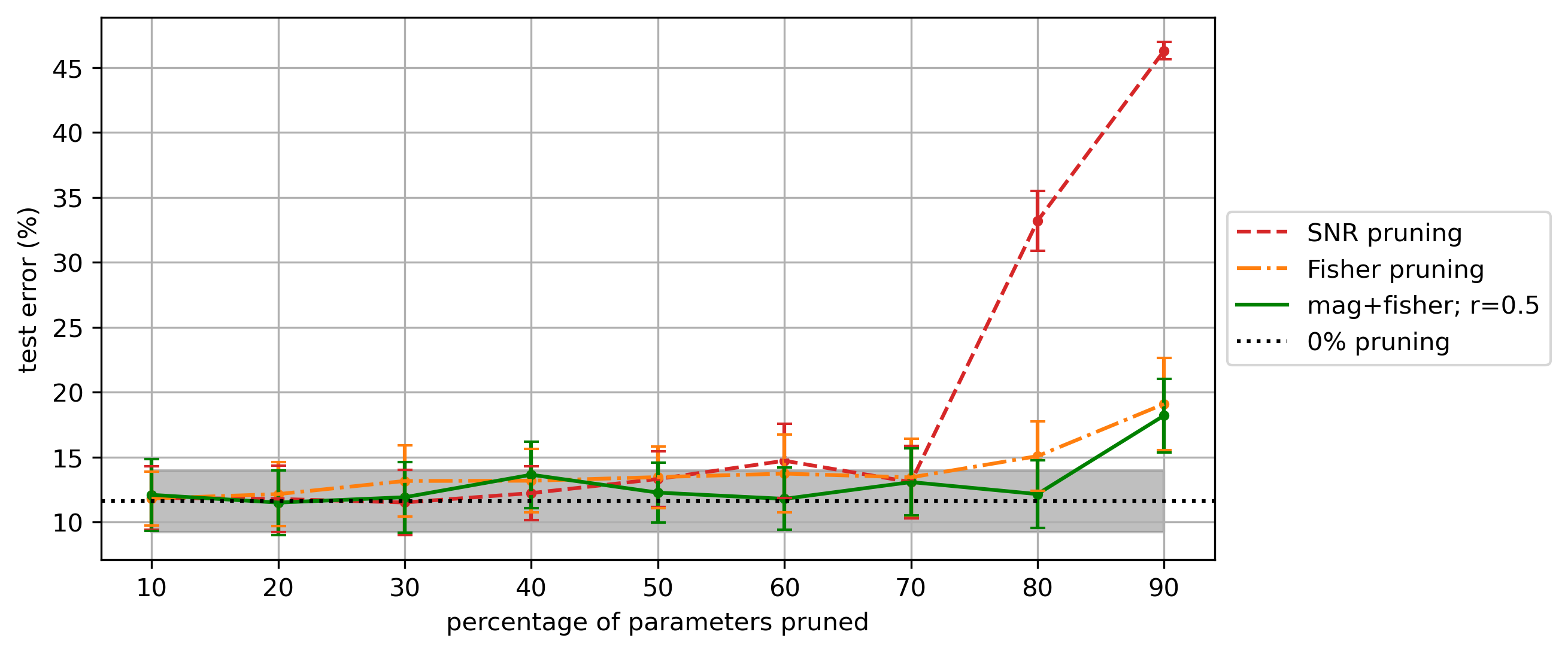}
    \caption[Comparison of model performance for different pruning methods]{Comparison of model performance for different pruning methods based on: SNR, Fisher information, a combination of magnitude and Fisher information. The grey shaded portion indicates the standard deviation of test error for the unpruned model.}
    \label{fig:fi_pruning}
\end{center}
\end{figure*}

The results of our pruning experiments are shown in Figure~\ref{fig:fi_pruning}. As also observed by \cite{tu2016reducing}, pruning the weights based on Fisher information alone does not allow for a large number of parameters to be pruned effectively because many values in the FIM diagonal are close to zero. We find this to be true for our model as well, and that only 10\% of the fully-connected layers can be pruned using Fisher information alone without incurring a significant penalty in performance. 

To remedy this, \cite{tu2016reducing} suggest combining Fisher pruning with magnitude-based pruning. Following their approach, we define a parameter, $r$, to determine the proportion of weights that are pruned by either of these methods. To prune $P$ parameters from the network, we perform the following steps in order: (i) remove the $P(1-r)$ weights with the lowest magnitude; (ii) remove the $P \, r$ parameters with the lowest FIM values. The parameter $r$ is between $(0,1)$ and is tuned like a hyper-parameter. We get an optimal pruning performance with $r = 0.5$. We find that up to 60\% of the fully-connected layer weights can be pruned using this method without a significant change in performance, which is double the volume of weights pruned by the SNR-based method discussed in Section~\ref{sec:pruning}. We refer to this method as Fisher pruning in the following sections for conciseness.

\subsubsection{Analysis of Uncertainty Estimates for different pruning methods}

\begin{figure*}
\begin{center}
    \begin{subfigure}[b]{0.29\textwidth}
        \includegraphics[width=\textwidth]{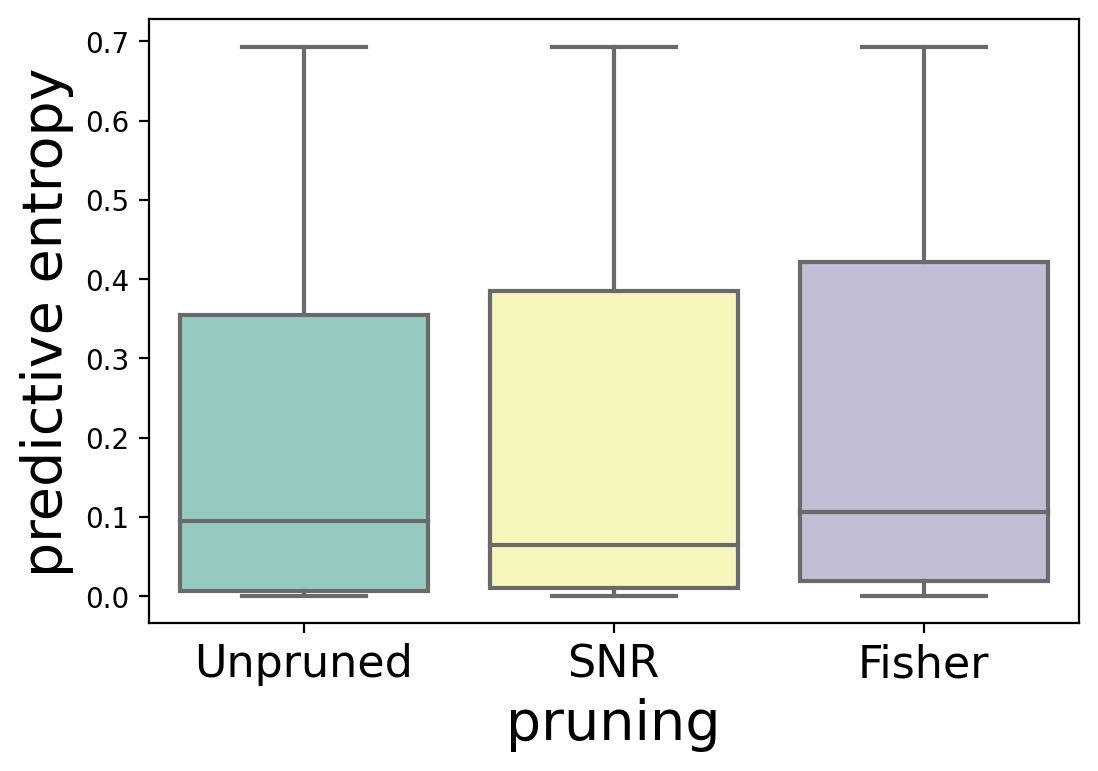}
        \caption[]{}
        \label{fig:mbconf_pruned_entropy}
    \end{subfigure} %
	\hfill
	\begin{subfigure}[b]{0.29\textwidth}
	    \includegraphics[width=\textwidth]{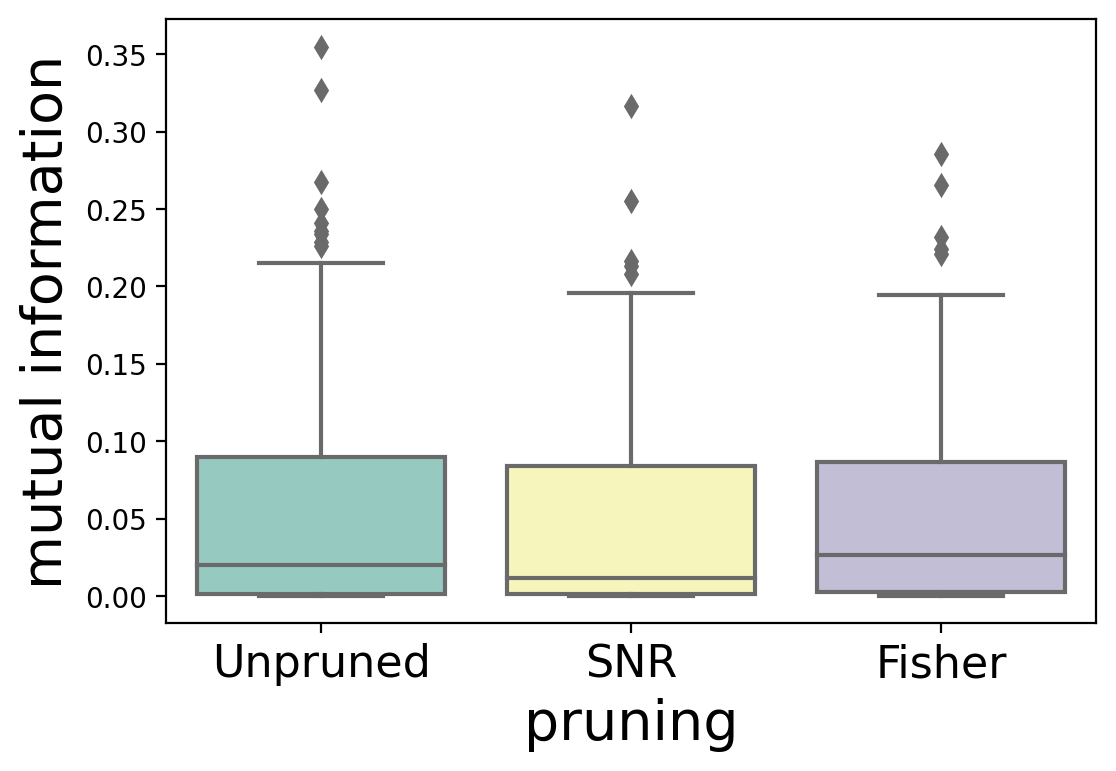}
        \caption[]{}
        \label{fig:mbconf_pruned_MI}
	\end{subfigure}
	\hfill
	\begin{subfigure}[b]{0.29\textwidth}
	    \includegraphics[width=\textwidth]{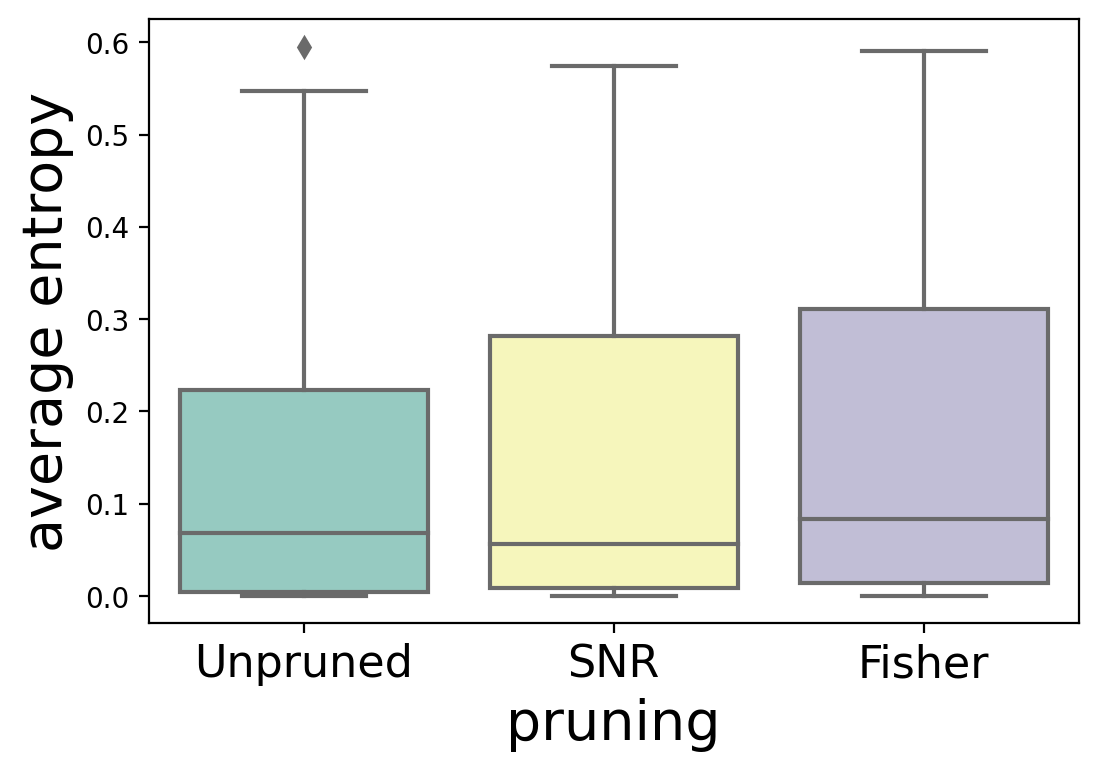}
        \caption[]{}
        \label{fig:mbconf_pruned_aleat}
	\end{subfigure}
	\caption[]{Distributions of uncertainty metrics for different pruning methods for the MiraBest Confident dataset. (a) Predictive uncertainty as measured using predictive entropy; (b) Epistemic uncertainty as measured using mutual information; and (c) Aleatoric uncertainty as measured using average entropy. For a fuller explanation of these metrics, please see Section~\ref{sec:uncertainty}.}
	\label{fig:mbconf_pruned_uncert} 
\end{center}
\end{figure*}

The effect of pruning on uncertainty quantification for the MiraBest Confident test set is shown in Figure~\ref{fig:mbconf_pruned_uncert}. We plot uncertainty metrics for the two pruning methods discussed in this work: (i) based on SNR, with 30\% pruning, and (ii) Fisher pruning which is based on magnitude combined with Fisher information, with 60\% pruning, and compare them to the metrics obtained for the unpruned model. We complement this with an analysis of the change in uncertainty calibration error for the two pruning methods considered in this work.

\begin{figure*}
\begin{center}
    \begin{subfigure}[b]{0.33\textwidth}
        \includegraphics[width=\textwidth]{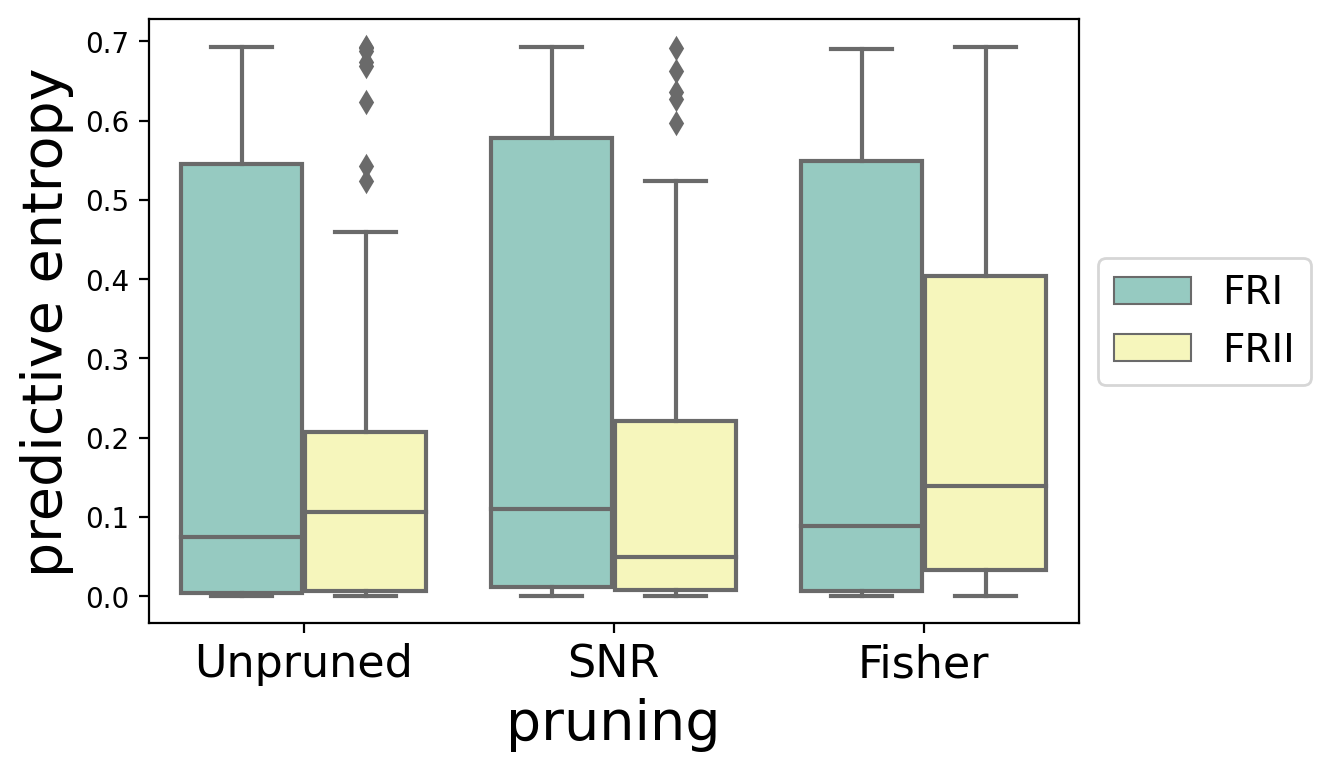}
        \caption[]{}
        \label{fig:fr_mbconf_pruned_entropy}
    \end{subfigure} %
	\hfill
	\begin{subfigure}[b]{0.33\textwidth}
	    \includegraphics[width=\textwidth]{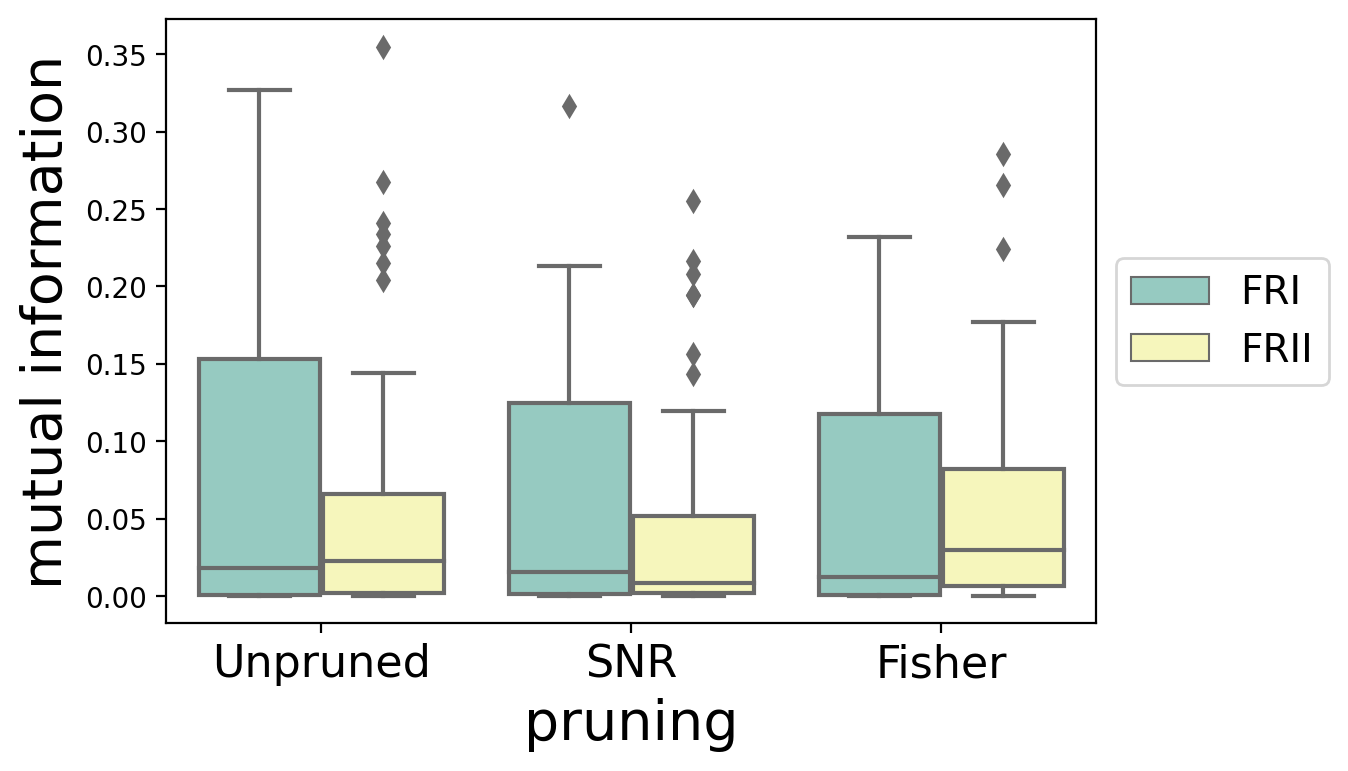}
        \caption[]{}
        \label{fig:fr_mbconf_pruned_MI}
	\end{subfigure}
	\hfill
	\begin{subfigure}[b]{0.33\textwidth}
	    \includegraphics[width=\textwidth]{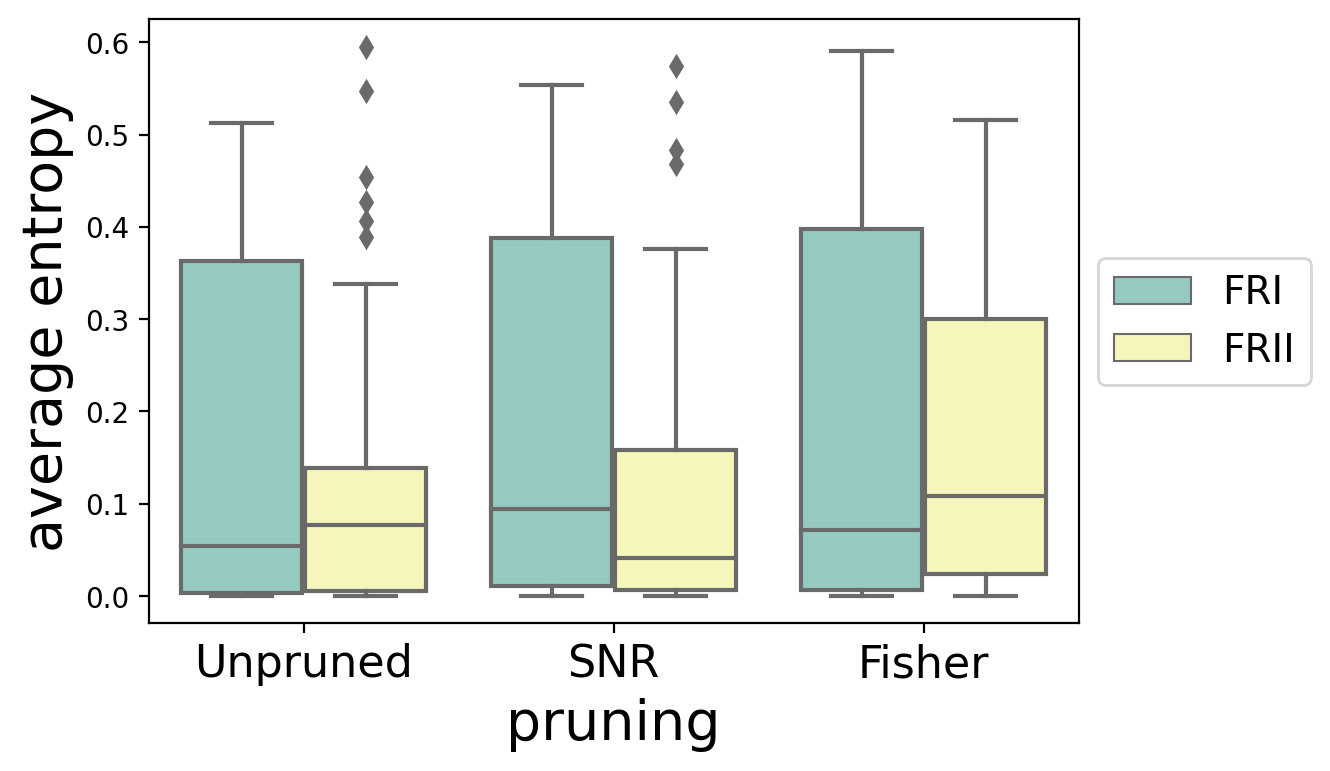}
        \caption[]{}
        \label{fig:fr_mbconf_pruned_aleat}
	\end{subfigure}
	\caption[]{Class-wise distributions of uncertainty metrics for different pruning methods for the MiraBest Confident dataset. (a) Predictive uncertainty as measured using predictive entropy; (b) Epistemic uncertainty as measured using mutual information; and (c) Aleatoric uncertainty as measured using average entropy. For a fuller explanation of these metrics, please see Section~\ref{sec:uncertainty}.}
	\label{fig:fr_mbconf_pruned_uncert} 
\end{center}
\end{figure*}

From Figure~\ref{fig:mbconf_pruned_entropy} we note that predictive entropy increases with pruning. The interquartile range increases for both pruning methods. This increase is mainly due to an increase in predictive entropy for FRI galaxies in case of SNR pruning and for FRII galaxies in case of Fisher pruning, see Figure~\ref{fig:fr_mbconf_pruned_entropy}. The median predictive entropy and interquartile range increase for FRIIs with Fisher pruning.

We also find that the classwise uncertainty calibration error (cUCE) of predictive entropy increases with pruning, and that this increase is larger in the case of SNR pruning, see Table~\ref{tab:calibration}.
SNR pruning has more of an adverse effect on the classification of FRI samples than FRII samples, whereas the effect of Fisher pruning is similar for both classes. 

Pruning does not seem to have a large effect on the distributions of mutual information, which narrow by a small amount for both pruning methods, see Figure~\ref{fig:mbconf_pruned_MI}. The classwise uncertainty calibration error does not increase significantly with SNR pruning and reduces with Fisher pruning, see Table~\ref{tab:calibration}.

\begin{table}
\centering
\caption[Uncertainty Calibration Error]{Percentage classwise Uncertainty Calibration Error (cUCE) on MiraBest Confident test set for our BBB-CNN model trained with a Laplace prior, pruned to its threshold limit for SNR and Fisher pruning. The percentage cUCE is shown separately for the predictive entropy (PE), mutual information (MI) and average entropy (AE) as calculated on the MiraBest Confident test set.}
	\begin{tabular}{llccc}
		 &  & \multicolumn{3}{c}{\textbf{\% cUCE}} \\\cmidrule{3-5}
    \textbf{Prior} & \textbf{Pruning}  & \textbf{PE} & \textbf{MI}& \textbf{AE} \\
    \hline
    Laplace & Unpruned & 9.69  &  16.37 & 10.84\\
        & SNR & 14.35 & 16.82 & 13.93  \\
        & Fisher & 13.43 & 15.29 & 11.25\\ 
   	\end{tabular}
   \label{tab:calibration}
\end{table}

Looking at the class-wise distributions of mutual information in Figure~\ref{fig:fr_mbconf_pruned_MI}, we can see that both pruning methods narrow the distribution of mutual information for FRI galaxies. However, SNR pruning increases the uncertainty calibration error for FRIs and decreases UCE for FRIIs. Fisher pruning on the other hand decreases UCE for FRIs and increases UCE for FRIIs. Therefore, SNR and Fisher pruning both seem to be adversely affecting the uncertainty calibration for one of the two classes. 

The average entropy increases with both pruning methods and the distribution becomes more broad in the case of Fisher pruning, see Figure~\ref{fig:mbconf_pruned_aleat}. However, we also find that Fisher pruning does not significantly change cUCE and that there is more increase in cUCE with SNR pruning, see Table~\ref{tab:calibration}. 

Figure~\ref{fig:fr_mbconf_pruned_aleat} shows that average entropy increases with pruning for FRI galaxies for both pruning methods and there is a greater increase in average entropy with Fisher pruning for FRII galaxies. We find that SNR pruning increases UCE for both FRI and FRII galaxies, but the increase is more significant for FRI galaxies. Fisher pruning reduces UCE for FRIs and increases UCE for FRIIs.

Whilst we have described the differing effect of alternative pruning methods on a class-wise basis, we note that pruning itself cannot be applied selectively by class since it depends on the model parameters. However, based on the fundamental differences in the results from SNR and Fisher pruning, we suggest that  FRI galaxies are less influential on the gradients of the learnable parameters during training compared to FRIIs and that the learned model weights are less noisy for FRIIs compared to FRIs.

\cite{hullermeier2021} argue that epistemic and aleatoric uncertainty are mutable quantities as a function of model specification \citep[see also][]{kiureghian2009}; specifically, that the uncertainty contributed by each is affected by model complexity and class separability. For example, embedding a dataset into a higher dimensional feature space may result in greater separability of the target classes leading to lower aleatoric uncertainty, but the additional complexity from the higher dimensionality results in a model with higher epistemic uncertainty. In SNR-based pruning, we are reducing the dimensionality of the feature space, but at the same time we are also removing noisy weights (and hence noisy features) that may adversely affect class separability, and so it is expected that measures of both epistemic and aleatoric uncertainty will be changed as a function of pruning degree. 

\section{Cold posterior effect}
\label{sec:masegosa}

\begin{figure*}
    \centering
    \includegraphics[width=\textwidth]{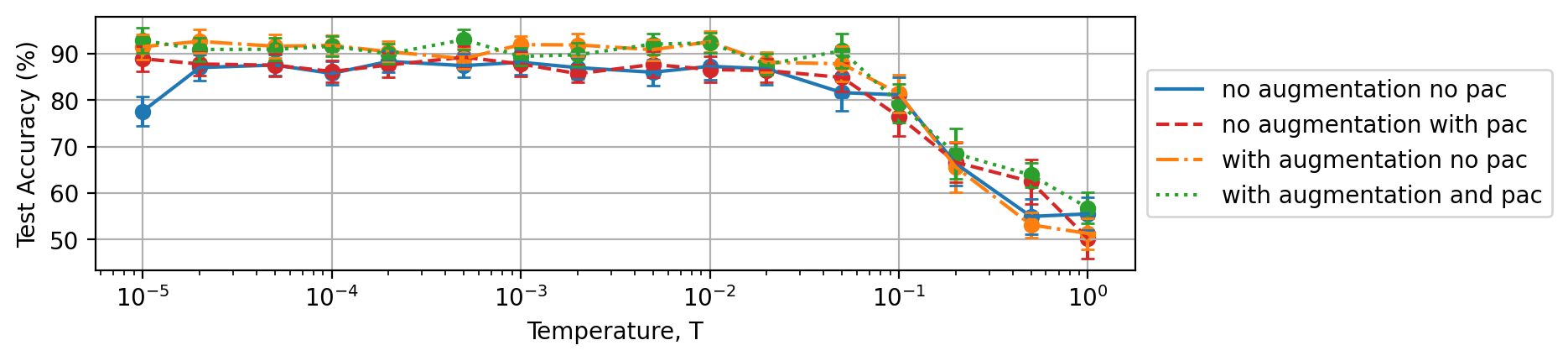}
    \caption{The “cold posterior” effect for the MiraBest classification problem, see Section~\ref{sec:masegosa} for details. Data are shown for the BBB model trained with a Laplace prior with no data augmentation and the original ELBO cost function (solid blue line), the BBB model with no data augmentation and the Masegosa posterior cost function (red dashed line), the BBB model with data augmentation and the original ELBO function (orange dot-dash line), and the BBB model with data augmentation and the Masegosa posterior cost function (green dotted line).}
    \label{fig:cold}
\end{figure*}

In Section~\ref{sec:cold} we found that we can improve the generalization performance of our BBB model significantly by cooling the posterior with a temperature $T\ll1$, deviating from the true Bayes posterior. This cold posterior effect is shown in detail in Figure~\ref{fig:cold} for our model trained on radio galaxies with the Laplace prior. In Section~\ref{sec:cold} we modified our cost function to down-weight the posterior in Equation~\ref{eq:tempered_cost_blundell} using a temperature term, $T$, with all subsequent experiments performed using $T=10^{-2}$. In Figure~\ref{fig:cold} we show the effect of varying $T$ over a wide range of values.

These results suggest that some component of the Bayesian framework in the context of this application is misspecified and it becomes difficult to justify using a Bayesian approach to these models whilst artificially reducing the effect of the components that make the learning Bayesian in the first place.

Finding an explanation for the cold posterior effect is an active area of research and several hypotheses have been proposed to explain this effect \citep{wenzel2020}: use of uninformative priors, such as the standard Gaussian, which may lead to prior misspecification \citep{fortuin2021priors}; model misspecification; data augmentation or dataset curation issues which lead to likelihood misspecification \citep{aitchison2021a, nabarro2021data}. Here we consider two approaches for investigating the effect of these misspecifications. 

\subsection{Model Misspecification}

We examine whether the cold posterior effect in our results is due to model misspecification by optimising the model with a modified cost function, following the work of \citet{masegosa2019learning}. The cost function is modified on the basis of PAC-Bayesian theory. PAC (Probably Approximately Correct) theory has its roots in Statistical Learning Theory and was first described in \citet{valiant1984theory} as a method for evaluating learnability, i.e. how well a machine learns hypotheses given a set of examples. Although PAC started out as a frequentist framework, it was soon combined with Bayesian principles. It has now evolved into a formalised mathematical theory used to give statistical guarantees on the performance of machine learning algorithms by placing bounds on their generalisation performance.

According to the PAC theory, we can obtain an approximately correct upper bound on the generalisation performance of a model, as measured by test loss for example, which holds true with an arbitrarily high probability as more data is collected, hence the name "Probably Approximately Correct". Since the goal of any learning algorithm is to minimise the generalisation gap, which is defined as the difference between the out-of-sample or theoretical loss, $L(\theta)$, and the in-sample or empirical loss, $\hat{L}(\theta)$, PAC inequalities can be used to define new learning strategies to train models. Using PAC theory, we can obtain the following inequality:
\begin{equation}
    L(\theta) \leq \hat{L}(\theta) + \epsilon(\delta, D) ,
\end{equation}
where $\delta$ is a confidence parameter that defines the probability that a sample in the training set is misleading, and $\epsilon(\delta, D_{{\rm train}})$ is an upper bound on the generalisation gap.

\citet{mcallester1999some} presented PAC-Bayesian inequalities which combine PAC-learning with Bayesian principles and provide guarantees on the performance of \emph{generalised} Bayesian algorithms. These algorithms are referred to as generalised because the PAC-Bayes framework has similar components to the Bayesian framework: a prior, $\pi$, defined over a set of hypotheses, $\theta \in \Theta$, and a posterior, $\rho$, which is updated using Bayes-rule style updates using samples from a data generating distribution, $\nu(x)$. But these bounds hold true for all choices of priors, whereas there is no guarantee on performance in Bayesian inference if the dataset is not generated from the prior distribution i.e. if the prior assumptions are incorrect. The bounds also hold true for all choices of posteriors, so in principle we can have model-free learning. However, traditionally most of the PAC-Bayes bounds are only applicable to bounded loss functions and this has made it difficult to apply them to the unbounded loss functions which are typically used to train neural networks. Fortunately more recent works have introduced PAC Bayes bounds for unbounded losses as well \citep[e.g.][]{alquier2016properties, germain2016pac, shalaeva2020improved}. We refer the reader to \citet{guedj2019primer} for an overview of the PAC-Bayesian framework.

Learning in the PAC-Bayesian framework happens such that an optimal value of the posterior, $\rho^*$, is found that minimises the KL divergence between the data generating distribution, $\nu(x)$, and the posterior predictive distribution given by $ \mathbb{E}_{\rho} [p(x|\theta)]$:
\begin{equation}
    \rho^* = \operatorname*{arg\,min}_{\rho} {\rm KL}[\nu(x) ~ || ~ \mathbb{E}_{\rho} [p(x|\theta)] ~ ] .
\end{equation}

Minimising this KL divergence is equivalent to minimising the cross-entropy (CE) function,
\begin{equation}
    \rho^* = \operatorname*{arg\,min}_{\rho} {\rm CE} (\rho), 
\end{equation}
which is the expected log loss of the posterior predictive distribution, $p(x|\theta)$,  with respect to the data generating distribution, $\nu(x)$:
\begin{equation}
    {\rm CE}(\rho) = \mathbb{E}_{\nu(x)} [-\log \mathbb{E}_{\rho(\theta)}[p(x|\theta)]] .
\end{equation}
Thus, by minimising this CE function, we can find the optimal $\rho^*$. This cross entropy loss is bounded by the expected log loss of the posterior predictive distribution as:
\begin{equation}
    CE(\rho) \leq \mathbb{E}_{\rho(\theta)}[L(\theta)]. 
    \label{eq:ce_bound}
\end{equation}
This is an example of what is known as an
`oracle' bound, since the inequality depends on the unknown data generating distribution, $\nu(x)$. It is also an example of a first order Jensen inequality, which gives a linear bound such that:
\begin{equation}
    \mathbb{E}_{\nu(x)} [-\log \mathbb{E}_{\rho(\theta)}[p(x|\theta)]] \leq \mathbb{E}_{\rho(\theta)}[\mathbb{E}_{\nu(x)}[-\log p(x|\theta)]] ,
\end{equation}
which is an expansion of the terms given in Equation~\ref{eq:ce_bound}.

\citet{germain2016pac} derived a first order PAC-Bayes bound for unbounded losses:
\begin{equation}
    {\rm CE}(\rho) \leq \mathbb{E}_{\rho(\theta)}[L(\theta)] \leq \mathbb{E}_{\rho(\theta)}[\hat{L}(\theta, D)] + \frac{{\rm KL}(\rho, \pi)}{c_1} + c_2, 
\end{equation}
where $c_1$ and $c_2$ are constants. In the same work, \citet{germain2016pac} also showed that under i.i.d. assumptions, the Bayesian posterior, $p(\theta|D)$, minimises this PAC-Bayes bound over the expected log loss, $\mathbb{E}_{\rho(\theta)}[L(\theta)]$, which bounds the cross-entropy loss.


Since variational inference is an approximation to the Bayesian marginal likelihood, we can use PAC-Bayesian bounds to optimise VI-based Bayesian NNs. By applying these bounds to train Bayesian neural networks, one deviates from the variational inference as defined in the Bayesian paradigm and moves towards a more generalised variational inference algorithm. Training a model with the modified cost function minimises an upper bound on the test loss and provides a more optimal learning strategy compared to optimising the Bayesian posterior or its approximations such as the ELBO function. To do this, the cost function is modified such that it minimises a second-order PAC-Bayesian bound on the cross entropy loss (CE), rather than the standard ELBO. Following the literature, we refer to this new objective function as a `Masegosa posterior'.

\subsubsection{Masegosa Posteriors}

\citet{masegosa2019learning} showed that the Bayesian posterior minimises a PAC-Bayes bound over the CE loss only when the model is perfectly specified. When the model is misspecified, the minimum of the CE loss is not equal to minimum of the expected log loss, and thus optimising the Bayesian posterior does not give an optimal learning strategy. Since this is more often the case, they propose an alternative posterior by introducing a \emph{variance term}, $\mathbb{V}(\rho)$, that measures the variance of the posterior predictive distribution. They define a second-order oracle and Jensen bound, which is given as:
\begin{equation}
    CE(\rho) \leq \mathbb{E}_{\rho(\theta)}[L(\theta)] - \mathbb{V}(\rho),
    \label{eq:second_oracle_bound}
\end{equation}
where
\begin{equation}
    \mathbb{V}(\rho) = \mathbb{E}_{\nu(x)} \left[ \frac{1}{2 {\rm max}_\theta p(x|\theta)^2} \mathbb{E}_{\rho(\theta)}[(p(x|\theta) - p(x))^2 ]  \right]
    \label{eq:variance_term}
\end{equation}
is the variance term.

Since the true data generating distribution, $\nu(x)$, is not known, the authors place an upper bound on Equation~\ref{eq:second_oracle_bound} using a second-order PAC-Bayes bound:

\begin{equation}
    CE(\rho) \leq \mathbb{E}_{\rho(\theta)}[L(\theta] - \mathbb{V}(\rho) \leq \mathbb{E}_{\rho(\theta)}[\hat{L}(\theta,D)] - \mathbb{\hat{V}}(\rho, D) +  \frac{{\rm KL}}{c_1} + c_2
    \label{eq:pac2bayes}
\end{equation}

This alternative posterior is compatible with VI and we can modify our cost function to test the hypothesis that the cold posterior effect observed in our work is due to model misspecification. Instead of optimising the ELBO function in Equation~\ref{eq:cost}, we optimise the following function:
\begin{equation}
    \operatorname*{arg\,min}_{\theta} {\rm KL} [q(w|\theta)|P(w)] - \mathbb{E}_{q(w|\theta)} [\log P(D|w)] - \mathbb{\hat{V}}(q|D) .
\end{equation}

The empirical variance term, $\mathbb{\hat{V}}$, can be numerically calculated as follows:
\begin{eqnarray}
\nonumber    \mathbb{\hat{V}}(w, w\:', D) &=&  \exp(2 \log P(D|w )- 2 m_{D})\\
\nonumber    && - \exp( \log P(D|w ) + \log P(D|w\:' ) - 2 m_{D}),\\
    \label{eq:empirical_variance_term}
\end{eqnarray}
where $w, w\:'$ are samples from the variational posterior, $q(w|\theta)$, $D$ is the training data and $m_{D}$ is given as:
\begin{equation}
    m_{D} = \operatorname*{max}_{w} \log P(D|w).
\end{equation}

The results of this modification for a range of temperatures is shown in Figure~\ref{fig:cold}, where it can be seen that the Masegosa posterior PAC bound (red dashed line) improves the cold posterior effect over the original BBB model (solid blue line) slightly, but does not fully compensate for the overall behaviour. 

\subsection{Likelihood Misspecification}
\label{sec:augmentation}

\cite{aitchison2021a} suggested that rather than prior or model misspecification, the cold posterior effect might be caused by over-curation of training datasets leading to likelihood misspecification, where the training data was not statistically representative of the underlying data distribution. They showed that for highly curated datasets, such as CIFAR-10, the cold posterior effect could be mitigated by adding label noise to the training data. Other works in this area have suggested that unprincipled data augmentation could be a contributing factor to the cold posterior effect \citep[e.g.][]{nabarro2021data, izmailov21}. 

The MiraBest dataset used in this work is likely to be similarly subject to some over-curation in the same sense as CIFAR-10 as it was compiled using an average or consensus labelling scheme from multiple human classifiers. For CIFAR-10, \cite{aitchison2021a} introduced label noise by augmenting the original dataset using all individual classifications from fifty human classifiers in their work, which were provided by the CIFAR-10H dataset \citep{peterson2019}. Here we do not have access to the individual classifications for the MiraBest dataset, but we are able to augment our dataset in a more standard manner using rotations. 

We find that the cold posterior effect observed in our work \emph{reduces} slightly with data augmentation, see Figure~\ref{fig:cold}. We suggest that this is because we have augmented the MiraBest dataset using principled methods that correspond to an informed prior for how radio galaxies are oriented, as radio galaxy class is assumed to be equivariant to orientation and chirality  \citep[see e.g.][]{ntwaetsile2021,e2cnn}. Figure~\ref{fig:cold} shows the cold posterior effect for the radio galaxy classification problem addressed in this work both with (orange dot-dash line) and without (blue solid line) data augmentation. It can be seen that there is an improvement in performance at temperatures below $T=0.01$ causing the test accuracy to reach a plateau at higher temperatures than for unaugmented data. Data augmentation also improves performance at temperatures above $T=0.01$, however we also find that the uncertainty calibration error increases for the model trained with augmented data for $T = 0.01$.




Figure~\ref{fig:cold} also shows that combining a Masegosa posterior with data augmentation provides the most significant improvement to the cold posterior effect (green dotted line); however it does not rectify it completely. Since the Masegosa posterior is a more complete test of model misspecification than the data augmentation used here is of likelihood misspecification, we suggest that a key element for exploring this problem in future may be the availability of radio astronomy training sets that do not only present average or consensus target labels, but instead include all individual labels from human classifiers. 



\section{Conclusions}
\label{sec:conclusion}

In this work we have presented the first application of a variational inference based approach to deep learning classification of radio galaxies, using a binary FRI/FRII classification. Using a Bayesian Convolutional Neural Network based on the Bayes by Backprop (BBB) algorithm, we have shown that posterior uncertainties on the predictions of the model can be estimated by making a variational approximation to the posterior probability distribution over the model parameters.

We have considered the use of four different prior distributions over the parameters of our model. 
We find that a model trained with a Laplace prior performs better than one using a Gaussian prior in terms of mean test error by $\sim 3\%$, and better than a GMM prior by $\sim 1\%$; a model trained with a Laplace Mixture Model prior performs worse than those using Gaussian priors and the 
Laplace prior. We also find that the calibration of the posterior uncertainties for a model trained using a Laplace prior is better than for models trained using the other priors considered in this work. This suggests that learning in this case may benefit from sparser weights. However, given the uncertainties on these values we cannot yet draw statistically significant conclusions for prior selection. We also note that this work uses relatively simple priors and that future extensions will look more closely at prior specification and whether more informative priors can help learning.

We note that we obtain a larger error value than that obtained by other neural network based models trained on the MiraBest Confident dataset, but emphasise that other works have used data augmentation to increase the size of the dataset, whereas we have only used the original samples. This allowed us to study the use of VI-based neural networks on small datasets as well. If we include data augmentation we obtain a comparable performance to previously published results; however this comes at the cost of increased uncertainty calibration error. Thus there is a trade-off between standard models, which are somewhat more accurate, and BBB, which is reasonably accurate while giving more reliable posteriors and therefore potentially more scientifically useful.

Our analysis of different measures of uncertainty for our deep learning model indicates that model uncertainty is correlated with the degree of belief of the human classifiers who originally assigned the labels in the MiraBest dataset. We find that our BBB model trained on confidently labelled radio galaxies is able to reliably estimate its confidence in predictions when presented with radio galaxies that have been classified with a lower degree of confidence. Notably, all measures of uncertainty are higher for the samples qualified as \emph{Uncertain}. The model also made predictions with higher uncertainty for a sample of \emph{Hybrid} radio galaxies, which was expected as these samples were not present in the training data, but contained FRI/FRII like components nevertheless. Looking more closely at the class-wise distributions of uncertainties, we found that FRII type objects are associated with a lower degree of uncertainty than FRIs. Among the classes of the Hybrid samples, we found that the uncertainty was higher for the confidently labelled samples compared to the uncertainly labelled Hybrid samples. We suggest that this may be because objects with uncertain labels are more similar to the FRI/FRII samples the model was trained on, which is why the human classifiers were uncertain in their classification as a Hybrid.


We have explored different weight pruning approaches with the motivation of reducing the storage and computation cost of these models at deployment. We find that using a SNR based method using posterior means and variances allows the fully-connected layers of the model to be pruned by up to 30\%, but a method that combines Fisher information with weight magnitudes allows an even higher proportion of weights to be pruned, by up to 60\%, without compromising the model performance. 
The effect of removing some of these weights can also be seen in the uncertainty metrics. We found that both the uncertainty and the uncertainty calibration error increase with model pruning. However, both pruning methods seem to affect FRI/FRII samples differently. Future work in this area could make a comparison of these methods with augmented data to verify whether one method should be preferred over the other. Another possible extension could be re-training a pruned model to test whether pruning improves the generalisation performance of a network.

Finally, we consider the cold posterior effect and its implications for the use of Bayesian deep learning with radio galaxy data in future. We find that the cold posterior effect is worse when using a GMM prior than for a Laplace prior, and we consider the hypothesis that further model misspecification may be causing the observed cold posterior effect. We test this hypothesis by retraining our model with a modified cost function that provides a loose PAC-Bayes bound over the cross-entropy loss. We find that although the modified cost function improves model performance slightly, it does not compensate for the cold posterior effect completely. 

We also consider the possibility of likelihood misspecification and test whether a principled data augmentation could improve the cold posterior effect. Similarly, we find that a small improvement is observed, but not a sufficiently large change to remove the effect entirely. Based on these results, we suggest that over-curation of the training dataset may be responsible for the majority of the cold posterior effect in radio galaxy classification and recommend that future labelling schemes for radio astronomy data retain full details of labelling from all human classifiers in order to test and potentially mitigate this effect more fully.


In this work we have considered a binary classification of morphology, but a diverse and complex population of galaxies exist in the radio universe. Understanding how populations of radio galaxies are distributed gives us insight into the effect of extrinsic and intrinsic factors that may have led to the morphologies, which in turn help shape our understanding of radio-loud AGN, their excitation and accretion modes, how they evolve and their relationship with their host galaxies and environments. Deep learning will play an important role in extracting scientific value from the next-generation of radio facilities and understanding how neural network models propagate uncertainties will be crucial for deploying these models scientifically.

\section*{Acknowledgements}

We thank the anonymous referee for their useful comments that improved this work. 
AMS, MW \& MB gratefully acknowledge support from Alan Turing Institute AI Fellowship EP/V030302/1. FP gratefully acknowledges support from STFC and IBM through the iCASE studentship ST/P006795/1. MB gratefully acknowledges support from the UK Science \& Technology Facilities Council (STFC).

\section*{Data Availability}
Code for this work is available at: \url{https://github.com/devinamhn/RadioGalaxies-BBB}. This work makes use of the MiraBest machine learning dataset, which is publically available under a Creative Commons 4.0 license at \url{https://doi.org/10.5281/zenodo.4288837}.



\bibliographystyle{mnras}
\bibliography{vi_paper} 








\bsp	
\label{lastpage}
\end{document}